\RequirePackage{ifpdf} 
\documentclass{PoS}

\usepackage[authoryear,square]{natbib}
\bibpunct{(}{)}{;}{a}{}{,}

\title{The SKA view of Gamma-ray Bursts}
\ShortTitle{GRB SKA}

\author{\speaker{Davide Burlon}$^{,1,2}$,
        Giancarlo Ghirlanda$^3$,  Alexander van der Horst$^4$, Tara Murphy$^{1,2}$, Ralph Wijers$^4$, Bryan Gaensler$^{1,2}$, Gabriele Ghisellini$^3$, Isabella Prandoni$^5$ \\
\llap{$^1$}Sydney Institute for Astronomy, The University of Sydney, NSW 2006, Australia,\,\\
\llap{$^2$}ARC Centre of Excellence for All-sky Astrophysics (CAASTRO),\,\\        
\llap{$^3$}INAF -- Osservatorio Astronomico di Brera, via E. Bianchi 46, I-23807 Merate (LC) - Italy,\,\\
\llap{$^4$}Anton Pannekoek Institute, University of Amsterdam, Science Park 904, 1098 XH Amsterdam, The Netherlands,\,\\
\llap{$^5$}INAF -- IRA, Via P. Gobetti 101, 40129 Bologna, Italy\\
Email:	\email{davide.burlon@sydney.edu.au}
}

\abstract{
Gamma-ray bursts (GRBs) are some of the most extreme events in the Universe. As well as providing a natural 
laboratory for investigating fundamental physical processes, they might trace the cosmic star 
formation rate up to extreme redshifts and probe the composition of the intergalactic medium over most of the Universe's history. 
Radio observations of GRBs play a key part in determining their physical properties, but currently they
are largely limited to follow-up observations of $\gamma$-ray-detected objects. 

The SKA will significantly increase our ability to study GRB afterglows, following up several hundred objects in the high frequency bands already in the ``early science'' implementation of the telescope. SKA1-MID Bands 4 (4~GHz) and 5 (9.2~GHz) will be particularly suited to the detection of these transient phenomena. 
The SKA will trace the peak of the emission, sampling the thick-to-thin
transition of the evolving spectrum, and follow-up the afterglow down to the time the ejecta slow down to 
non-relativistic speeds. 
The full SKA will be able to observe the afterglows across the non-relativistic transition, for $\sim 25$\% of the whole GRB population. 
This will allow us to get a significant insight into the true energy budget of GRBs, probe their surrounding density profile, and the shock microphysics. 
The SKA will also be able to routinely detect the elusive ``orphan afterglow'' emission, from the population of GRBs whose jets are not pointed towards the Earth. We expect that a deep all-sky survey such as SKA1-SUR will see around 300 orphan afterglows every week. We predict these detection to be $\gsim1000$ when the full SKA telescope will be operational.
}

\FullConference{
Advancing Astrophysics with the Square Kilometre Array\\
June 8-13, 2014\\
Giardini Naxos, Italy}

\newcommand{\skipthis}[1]{}
\newcommand{\sw}{\textit{Swift}}

\newcommand {\gsim}{ \lower .75ex \hbox{$\sim$} \llap{\raise .27ex \hbox{$>$}} } 
\newcommand {\lsim}{ \lower .75ex\hbox{$\sim$} \llap{\raise .27ex \hbox{$<$}} }

\def\eiso{$E_{\gamma, \rm iso}$}

\def\ekin{$E_{\rm kin, iso}$}
\def\ekinj{$E_{\rm kin, jet}$}

\def\th{$\theta_{\rm jet}$}

\def\thv{$\theta_{\rm view}$}

\newcommand\apj{ApJ}
\newcommand\aap{A\&A}
\newcommand\araa{ARA\&A}
\newcommand\aaps{A\&AS}
\newcommand\actaa{AcA}
\newcommand\mnras{MNRAS}
\newcommand\nat{Nature}
\newcommand\apjl{ApJL}
\newcommand\apjs{ApJS}

\newcommand\aj{AJ}

\newcommand\pasa{PASA}

\begin{document}

\section{Introduction}
Gamma-ray bursts (GRBs) are among the most extreme events in the Universe.
they provide a laboratory for investigating fundamental physical processes,
trace the cosmic star formation up to very high redshifts, and probe the
composition of the intergalactic medium over most of the Universe's
history.  Radio observations of GRBs play a key part in determining their
physical properties by complementing the broadband spectrum and providing
information unique to the radio regime. 

The origin of GRBs remained elusive for three decades after their initial discovery \citep{klebesadel1973}, but this changed with the discovery of their afterglows at X-ray \citep{costa1997}, optical \citep{vanparadijs1997}, and radio \citep{frail97} frequencies. 
GRBs occur at cosmological distances \citep{meegan1992,metzger1997}, and are generally thought to be caused by either the collapse of a massive star \citep{woosley1993}, or the binary merger of two neutron stars or a neutron star and a black hole \citep{eichler1989,narayan1992}. 
The former are the progenitors of so-called long GRBs, while the latter are associated with short GRBs. Long and short refer to the GRB duration in gamma-rays, and the divide between the two classes is at $\sim2$~s \citep{kouveliotou1993}, albeit it should be noted that the measured duration is energy-dependent. 
In both cases, the rapid accretion of material onto the compact object that is formed in these catastrophic events is thought to produce a pair of highly collimated relativistic jets, which decelerate and eventually evolve into a roughly isotropic blast wave. 
The isotropic equivalent kinetic energy released in a GRB is of order $10^{53-54}$ ergs, which is similar to that seen in supernova explosions when corrected for collimation and radiative efficiency \citep{vanparadijs2000}. 
However, GRB jets have bulk Lorentz factors of order of hundreds, and therefore relativistic beaming makes them some of the most luminous events in the Universe.

By studying the broadband emission of GRBs, from low-frequency radio waves to high-energy gamma-rays, we can determine the physical properties of the jet and its surroundings, and also the microphysics of the emission we observe. 
GRBs are amongst the highest redshiftobject known: the current highest spectroscopic redshift is $z=8.23$ \citep{tanvir2009}, and the photometric record-holder is $z=9.2$ \citep{cucchiara2011}.
Therefore they can be used as probes of the early Universe, of the host galaxies \cite[see e.g.][]{stanway14, perley13}, and of the medium in between their host galaxies and us. 
Furthermore, GRBs are excellent laboratories for studying particle acceleration physics, because they are not only accelerating electrons to extreme relativistic velocities to produce the observed emission, but are also prime candidates for producing ultra-high energy cosmic rays. 
In this chapter we first focus on the role of radio observations in GRB studies (Section~\ref{sec:obs}) and the progress that has been made in this field over the decade since the original SKA science case was written (Section~\ref{sec:progress}). 
We have performed simulations to give prospects for GRB studies in the SKA era, for GRBs whose jets point towards the Earth (Section~\ref{sect:pointing}), and also those which are pointed away and from which we can detect late-time emission when the GRB outflow enters our line of sight (Section~\ref{sect:orphan}). 
For a more comprehensive review on radio observations of GRBs and their role in GRB jet studies we refer to \citet{granot2014}.

\section{Radio observations of GRBs}\label{sec:obs}
Almost all GRBs have been detected first with $\gamma$-ray instruments and then followed up at other 
wavelengths \citep[see e.g. the interesting case reported by][]{cenko13}. 
Since the discovery of the first radio afterglow \citep{frail97} roughly one third of all GRBs with accurate locations has been detected at radio frequencies \citep{chandra12}. 
This rate is much lower than at higher observing frequencies, where for instance 93\% of GRBs detected in gamma-rays by the {\it Swift} satellite are also detected at X-ray frequencies and 75\% have been detected in optical. 
This difference is due to the fact that GRBs have radio flux densities typically at the sub-mJy level, which for a long time has been close to the sensitivity limits of even the largest radio observatories. 
Besides the sample of radio detected GRBs being sensitivity limited \citep{chandra12}, it has been claimed that there are in fact two populations of radio-bright and radio-faint GRBs \citep{hancock2013}. 
The latter has been found by stacking radio visibility data of many GRB follow-up observations. This could be tested with the upgraded Very Large Array and ultimately with the SKA. 
If there are indeed two populations of radio afterglows, the cause of this can be resolved by broadband modelling of a large sample of individual GRBs. 

\begin{figure}
\begin{center}
\includegraphics[viewport=0 0 410 783,clip,angle=-90,width=\textwidth]{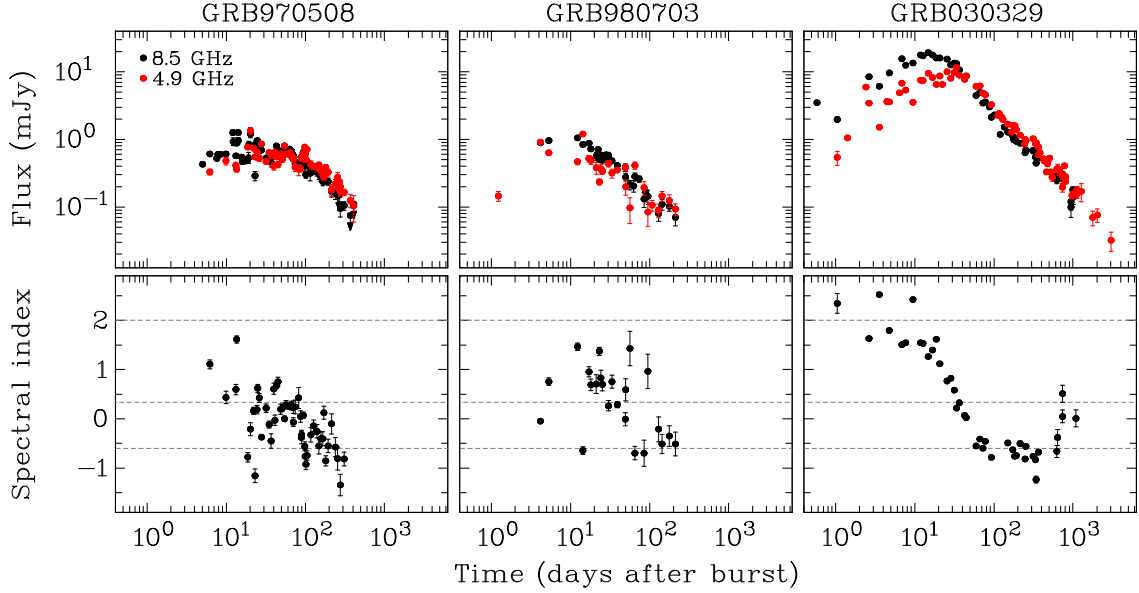}
\caption{Radio light curves at 4.9 and 8.5 GHz (top panels) and spectral indices 
(bottom panels) for the well-sampled, long-lasting GRBs 970508, 980703 and 030329 \citep{granot2014}. 
The spectral index $\alpha$ (where $F_{\nu}\propto\nu^{\alpha}$) between 4.9 and 8.5$\;$GHz 
varies significantly due to the spectral evolution and scintillation effects. 
The dashed lines in the bottom panels indicate spectral indices of $2$, $1/3$ and $-0.6$, which are expected below the self-absorption frequency, between the self-absorption and peak frequency, and above the peak frequency, respectively.}
\label{fig:lcspec}
\end{center}
\end{figure}

In the standard model for GRBs, the afterglow is thought to be emission from the blast wave at the front of the relativistic jet pointing towards us. 
As this shock sweeps up the ambient medium, it accelerates electrons (and protons) and amplifies the magnetic field, both necessary for the observed synchrotron radiation. 
Deceleration of the blast wave results in the peak of the spectrum moving from high to low observing frequencies over time \citep{sari1998}. 
This spectral peak is typically already below optical frequencies when the first observations commence, resulting in declining light curves at optical and X-ray frequencies, but rising light curves in the radio at early times, up to the peak occurring days or weeks after the initial gamma-ray trigger (see Figure~\ref{fig:lcspec}). 
Multi-frequency radio observations at later times can follow the evolution of both the peak frequency and peak flux. 
Furthermore, the radio regime is affected by synchrotron self-absorption, which makes early radio detections challenging, but when detected, time evolution of the self-absorption frequency provides important constraints on the blast wave physics. 
Therefore, the radio regime plays an important role in constructing the full broadband spectrum, to constrain both the macrophysics of the jet, i.e. the energetics and ambient medium density, and the microphysics of the electrons and magnetic fields necessary for synchrotron emission \citep{wijersg1999}. 

Radio afterglow emission peaks much later than at higher frequencies, and it also lasts much longer, for months or even years \citep[Figure~\ref{fig:lcspec}; e.g.,][]{frail2000,berger04,vanderhorst08}. 
Late-time observations can be used to study the GRB jet as it decelerates into the non-relativistic phase. 
This provides a unique probe for the dynamics of the jet, in particular for determining the true total kinetic energy in the blast wave; at late times the outflow is almost spherical and energy estimates are not complicated by relativistic effects (see Section~\ref{sect:nonrel} for a more detailed discussion). 
In addition to these late-time studies, the radio regime is also well suited for studying early-time phenomena which may go undetected at high frequencies, for instance emission from the reverse shock. 
When the blast wave at the front of the jet is formed, there is not only a shock moving forward into the ambient medium, but also a shock moving backwards into the ejecta \citep{sarip1999}. 
Under the right conditions this reverse shock will result in an optical flash in the first tens of seconds after the gamma-ray trigger \citep{akerlof1999}, which can only be detected if robotic telescopes are on target fast enough. 
However the peak of this emission moves to lower frequencies over time and can be probed at radio frequencies on a time-scale of hours to days \citep{kulkarni1999}. 
Multi-frequency observations of this reverse shock provide crucial information regarding the particle content of the jet and its internal magnetic field \citep[e.g.][and references therein]{perley14, anderson14}.

Another way in which radio observations play a key role in GRB studies is by providing evidence for the relativistic expansion of the jet. 
The source size and its evolution can be estimated either directly by means of Very Long Baseline Interferometry (VLBI), or indirectly by utilising the effects of interstellar scintillation. 
The former method has only been possible for one nearby GRB \citep{taylor2004}, and the source size measurements have been combined with its light curves to better constrain the physical parameters \citep{granot2005,mesler2013}.  
Indirect source size measurements are possible because of scintillation due to the local interstellar medium, which modulates the radio flux \citep{goodman1997}. 
While the angular size of the blast wave is initially smaller than the characteristic angular scale for interstellar scintillation, it grows with time until it eventually exceeds this scale and the modulations quench. 
This can be utilised in determining the source size and the expansion speed of the blast wave, which impose constraints on the modelling of GRB afterglows \citep{frail97}.
The future possibilities of VLBI observations of GRBs in the SKA era are discussed elsewhere in this book; see the Chapter by Paragi et al., ``Very Long Baseline Interferometry with the SKA'', in proceedings of ``Advancing Astrophysics with the Square Kilometre Array'', PoS(AASKA14)143.

All the observational characteristics at radio frequencies discussed so far are for GRBs in which the jet is oriented along our line of sight. 
Since the initial Lorentz factors of the shocks are of the order of a few hundred, relativistic beaming causes the initial $\gamma$-ray emission to go undetected if we are not observing a GRB within the jet opening angle. 
Therefore, we only detect a small fraction of the total number of events \citep{rhoads1997}. 
Estimates of collimation for individual GRBs are usually derived from the afterglow light curve: an achromatic break to a steeper decay indicates a transition across the jet opening angle. 
The time at which this `jet break' occurs, combined with the ambient medium density and the energetics of the GRB, leads to an estimate of the jet opening angle \citep{sariph1999}. 
This approach is limited by the difficulty in identifying an unambiguous jet break in the light curve and yields a wide variation of jet angles, but typically a few to a few tens of degrees. 
An alternative way to obtain a handle on jet opening angles, and on the total GRB rate whether beamed towards us or not, is to carry out a survey for GRB afterglows at very late times when the outflow is (quasi-)spherical and only detectable at radio frequencies. 
Since a large fraction of GRBs are associated with supernovae of Type Ib/c, it has been suggested to search for radio counterparts of relatively nearby supernovae of that type \citep{paczynski2001,granot2003}. 
Attempts to find such orphan radio afterglows have been unsuccessful thus far \citep{berger2003,soderberg2006,bietenholz2014}.

\section{Progress in the Last Decade}\label{sec:progress}
Our understanding of GRBs has increased significantly over the past decade, mainly because of improvements in observational capabilities across the electromagnetic spectrum. 
For instance, the {\it Swift} satellite has revealed very interesting X-ray behavior in the first minutes to hours \citep{nousek2006}, robotic telescopes have done the same at optical frequencies \citep[e.g.,][]{racusin2008}, and the {\it Fermi Gamma-ray Space Telescope} has opened the high-energy gamma-ray regime for GRBs, including emission at GeV energies \citep{ackermann2013}.
Furthermore, the GRB-supernova association has been further established for long GRBs \citep{woosley2006}, while afterglows of short GRBs have been discovered \citep{gehrels2005}, and strong indications for their connection with binary mergers of compact objects have been found \citep{tanvir2013}. 

From the radio perspective, GRB\,030329 had a major impact: the first high-luminosity GRB at low redshift, which also satisfied the GRB-supernova connection. 
It is the only GRB up to now for which the source size has been measured with VLBI, and the radio afterglow was so bright and long-lasting that it has been detected for almost a decade at low radio frequencies \citep{vanderhorst08,mesler2012}. 
The multi-frequency light curves have made it possible to follow the evolution of the blast wave from the ultra-relativistic to the non-relativistic phase and perform broadband modeling in the different phases. 
This has led to tight constraints on the physical parameters, and also a consistent picture together with modeling of the VLBI measurements \citep{granot2005,mesler2013}. 
The puzzle that still remains is that all the modeling work points to the ambient medium being homogeneous, and in that case a signature from the counter-jet (the jet originally pointed away from the Earth) should have been observed \citep{decolle2012}. This did not happen, questioning the existance and properties of this counter-jet. 

\begin{figure}
\begin{center}
\includegraphics[angle=-90,width=0.48\textwidth]{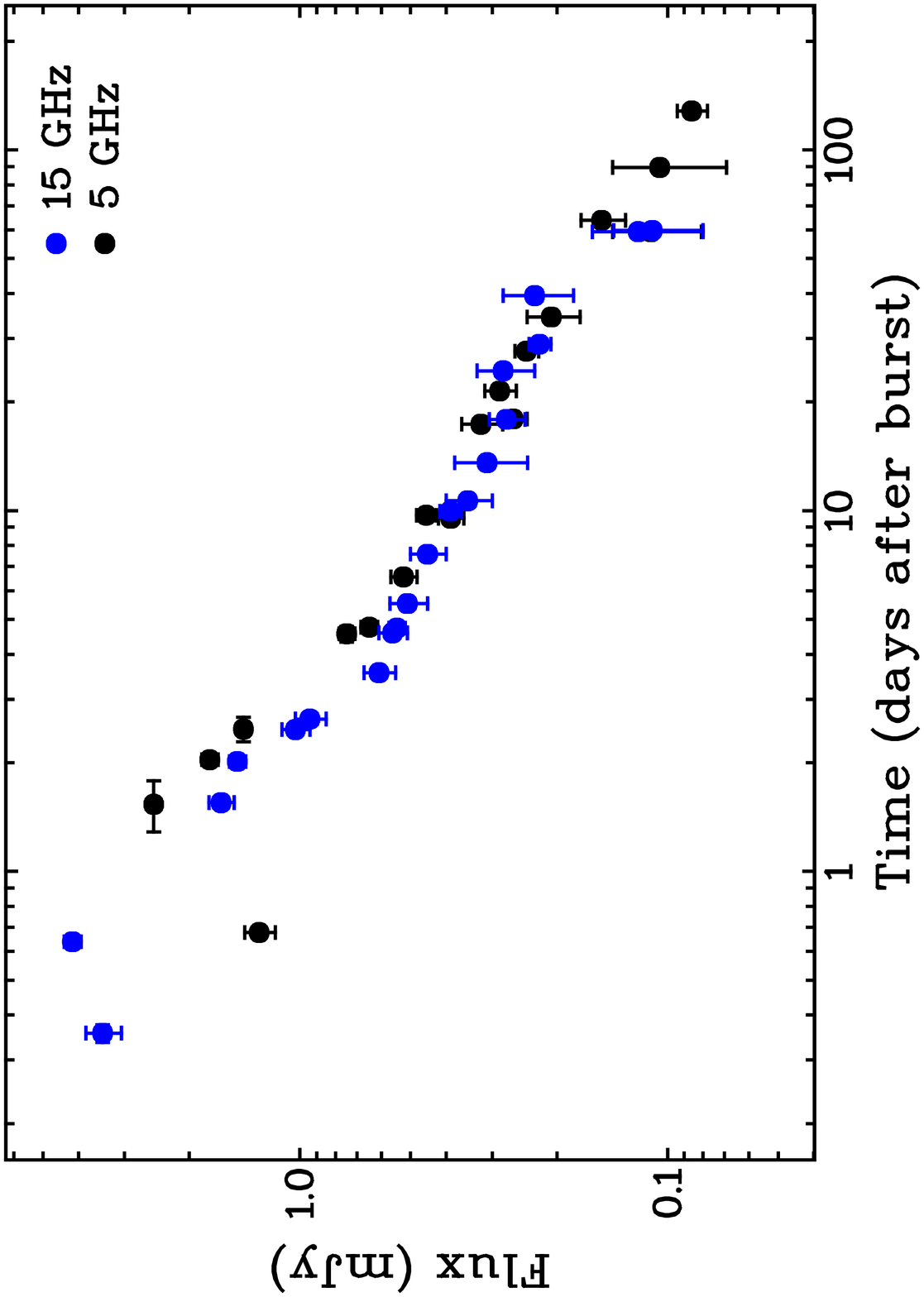}\hfill
\includegraphics[angle=-90,width=0.48\textwidth]{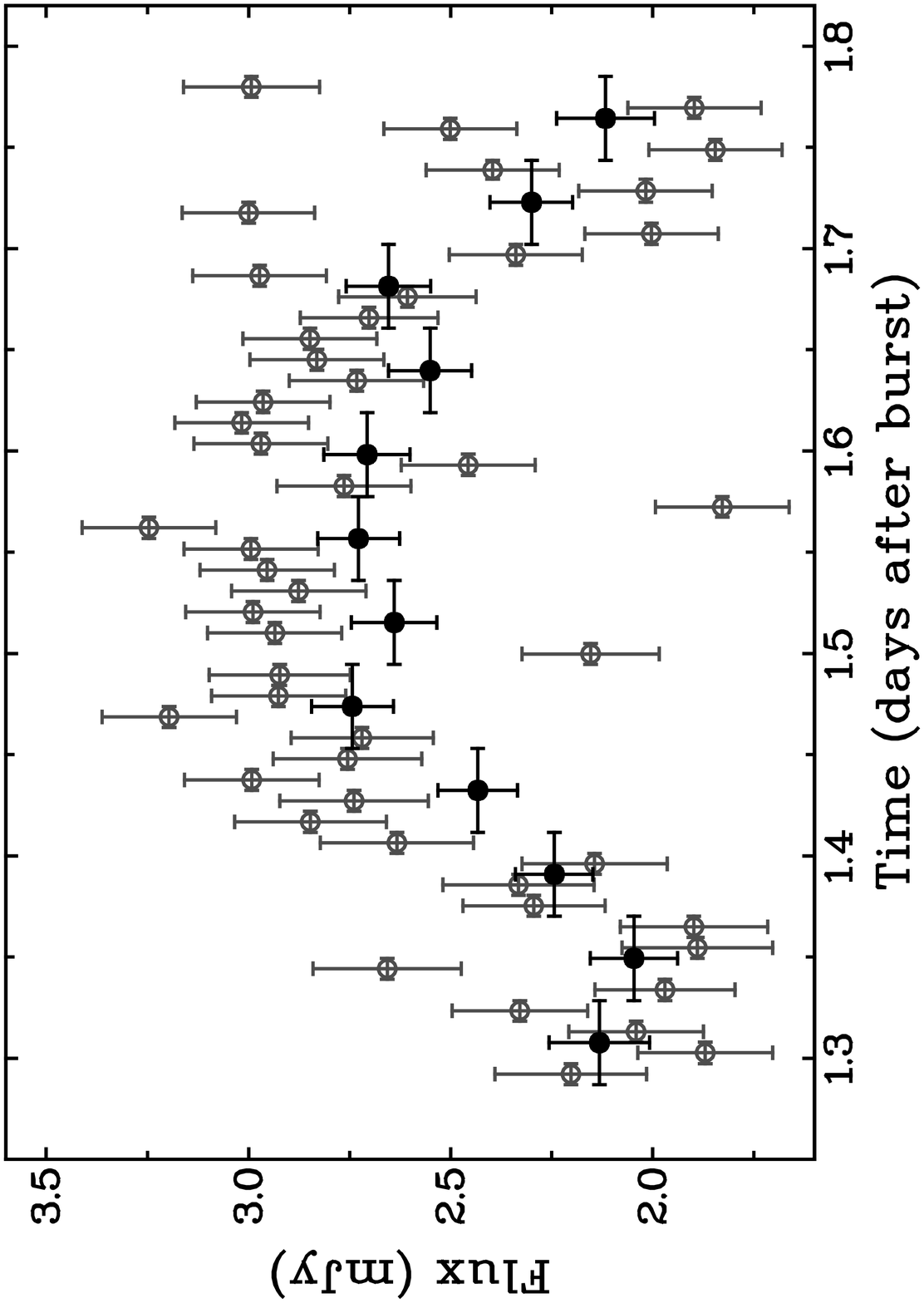}
\caption{The left panel \citep{anderson14} shows the radio light curves of the bright GRB\,130427A at 5 and 15 GHz \citep[see also][]{laskar2013, perley14, vanderhorst14}, showing the evolution of the radio peak moving from high to low frequencies at early times. 
The right panel shows strong variability at the peak of the 5 GHz light curve on a timescale of hours (solid symbols), and perhaps even shorter (15 minutes; open symbols), due to interstellar scintillation \citep{vanderhorst14}.}
\label{fig:130427A}
\end{center}
\end{figure}

Radio observations have played an important role in the broadband modeling of a larger sample of GRBs, e.g., very energetic GRBs detected by {\it Swift} \citep{cenko2010} and those for which high-energy gamma-rays have been detected by {\it Fermi} \citep{cenko2011}. 
These studies indicate that there is a large spread in the physical parameters, which has implications for models of GRB progenitors, particle acceleration, and magnetic field amplification by shocks \citep{granot2014}. 
A striking example is GRB\,130427A, another nearby, high-luminosity event with extremely good temporal and spectral coverage, over 10 orders of magnitude in time and 16 orders of magnitude in observing frequency \citep{ackermann2014,maselli2014}. 
A bright radio flare was observed in the first days, with the peak moving from high to low radio frequencies over time \citep[left hand panel of Figure~\ref{fig:130427A};][]{laskar2013, anderson14, perley14, vanderhorst14}. 
The earliest clear detection at 8~h and subsequent observations at high cadence have shown how early-time radio observations can contribute significantly to our understanding of the physics of both the forward and reverse shocks. 
This was also the second GRB for which radio variability on the timescale of hours has been reported \citep[right hand panel of Figure~\ref{fig:130427A};][]{vanderhorst14}, and in this case, as well as for GRB\,070125 \citep{chandra2008}, the variability strength and timescale are both consistent with interstellar scintillation effects. 

An important probe for getting a better understanding of the jet and magnetic field structure is polarization measurements. 
In the optical, linear polarization of the order of a few percent has been observed in emission from the forward shock \citep{covino1999,wijers1999}, while this can be tens of percent for the reverse shock \citep{mundell2013}. 
Optical circular polarization has recently been discovered, at a level of only a few tenths of a percent \citep{wiersema2014}. 
At radio frequencies the most stringent constraints have been obtained for GRB\,030329 at late times \citep{taylor2005}, and for GRB\,130427A at early times \citep{vanderhorst14}. 
The upper limits are only a few percent in both linear and circular polarization, and thus getting close to the optical detections. 
Deeper radio polarization measurements for bright GRBs, preferably at early times and for GRBs with optical polarization detections as well, will provide more insight into the magnetic field strength and structure inside the GRB jet.

Coherent radio emission has been predicted at earlier times than from the reverse shock, which peaks at a timescale of hours to days at radio frequencies. 
In these models the jet is magnetically dominated \citep{usov2000,sagiv2002,moortgat2006}, and detecting this emission would uniquely constrain the jet production and collimation mechanism. 
The first searches have not resulted in any firm detections \citep{bannister2012,obenberger2014}, but surveys with large field-of-view telescopes are vital for this research. 
Such studies should also clarify if (some of) the recently discovered fast radio bursts \citep{lorimer2007,thornton2013} are connected to GRBs \citep{zhang2014}. 

Radio surveys with sufficient sensitivity and sky coverage are not only important for prompt radio emission searches, but also for detecting orphan afterglows. 
Such studies would let us directly constrain the overall beaming fraction of GRBs, and yield the rate of the events that generate this phenomenon. 
Furthermore, since radio emission is unaffected by extinction, and a significant fraction of GRBs are thought to occur in highly obscured regions \citep{perley2009}, radio surveys can provide a new picture of the GRB population unbiased by their local environments. 
This ties in with studies of dark bursts, i.e. GRBs for which the optical afterglow is much dimmer than expected from the observed X-ray emission. 
These GRBs are either at a high redshift or there is significant optical extinction in the host galaxy. 
The latter is the most common explanation, and the amount of extinction can be estimated by modeling of the X-ray and radio light curves \citep{rol2007,zauderer2013}. 

\section{The GRB population pointing to Earth}\label{sect:pointing}
In this section we focus on GRBs that can be studied by the SKA after a trigger has been produced by $\gamma$-ray 
detectors, as now happens with \sw\ and \textit{Fermi} providing the high energy trigger and initiating the follow-up 
campaign. These are the GRBs that have their jets pointed towards the Earth. 

Substantial advances in breaking the degeneracies of the physical parameters of the afterglow emission process and, globally, of the GRB itself can only be possible through a considerable increase of the detection rate of GRBs coupled to the ability to follow the radio emission from early to very late times (i.e. at extremely faint flux densities). 

The key questions we will address are:
\begin{enumerate}
\item What are the prospects for the detection and follow up of GRBs by the SKA? 
\item What is the expected rate of GRBs that can be observed by the SKA? 
\item What is the best observational strategy, considering the wide frequency range covered by the SKA and the behaviour of the GRB emission at radio frequencies as a function of time? 
\end{enumerate}

\begin{figure}[!ht]
\includegraphics[width=7.3cm,trim=40 10 20 10]{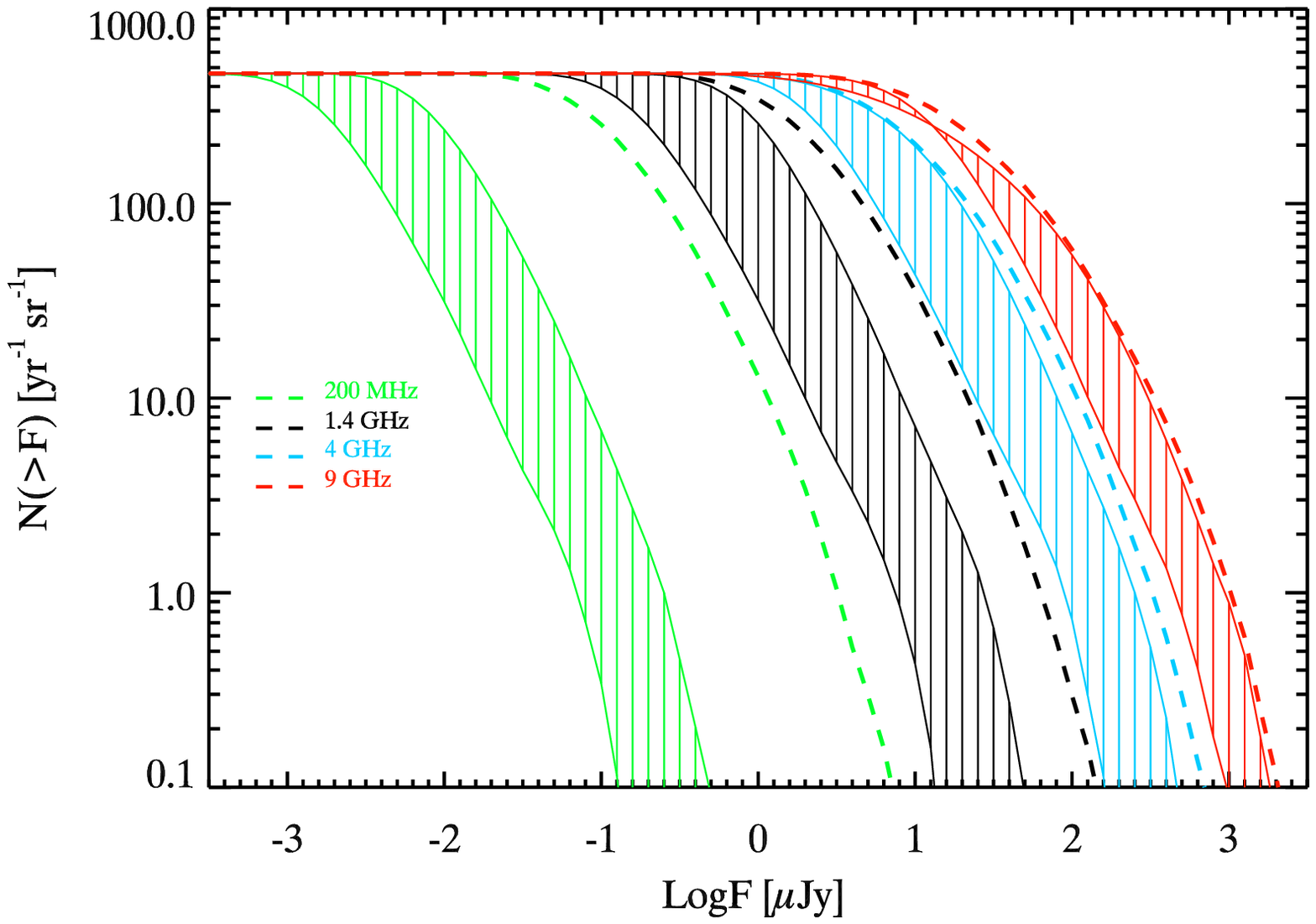}
\hspace{0.5cm }
\includegraphics[width=7.3cm,trim=40 10 20 10]{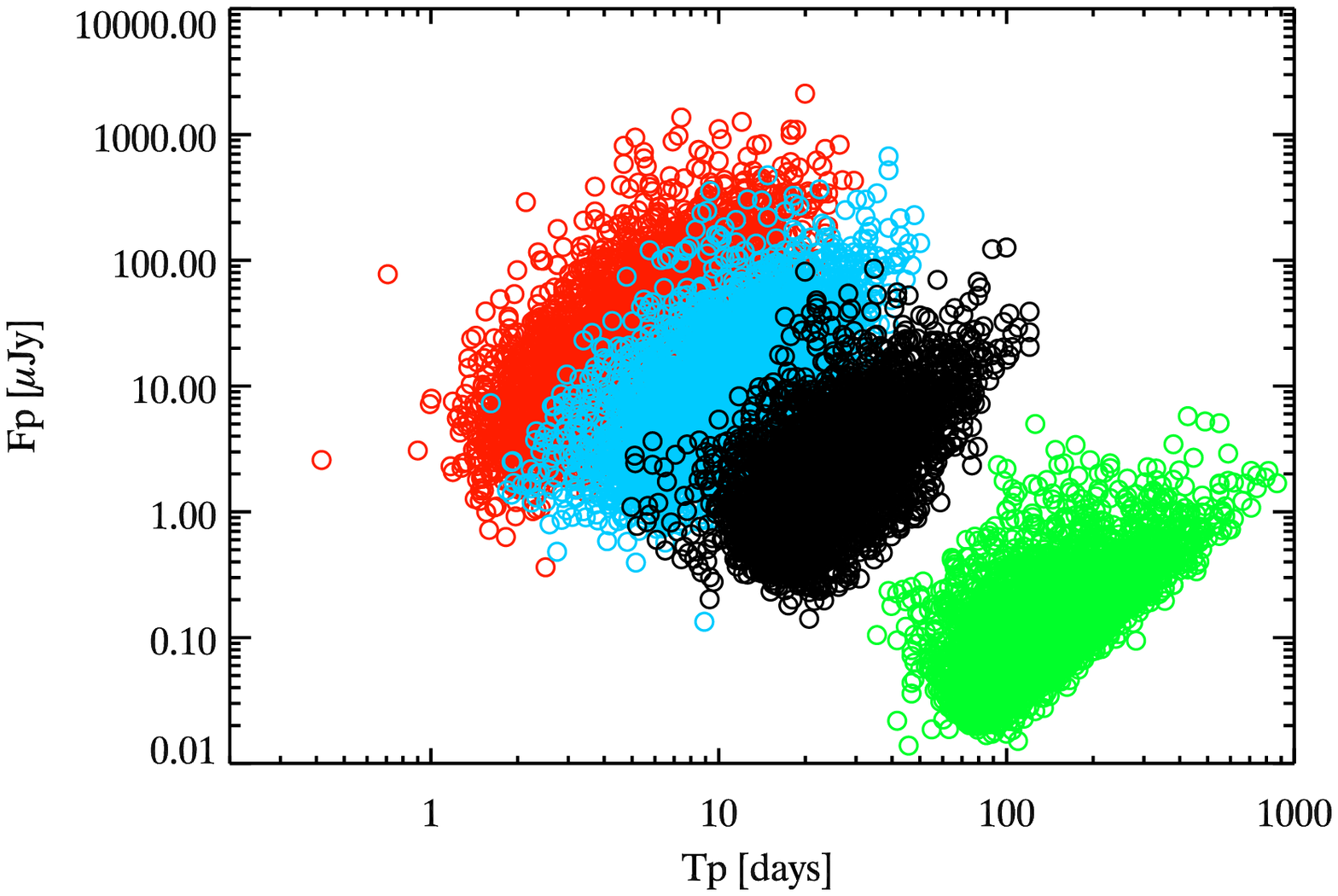}
\caption{\textit{Left}: Cumulative flux distribution of afterglows at SKA1-LOW and SKA1-MID observing bands 2, 4, and 5 (from left to right). The shaded area represents the span between 2 and 10 days (left and right boundary respectively). The dashed line represent the flux at the peak time. 
\textit{Right:} 
Afterglow peak flux vs. observed time of the peak (which happens at the time the peak of the spectrum crosses the observed band) for different SKA frequencies. The colours (red, blue, black, and green) represent bands 5, 4, 2 of SKA1-MID and SKA1-LOW, i.e. 9.2, 4, 1.4 and 0.2~GHz respectively.}\label{logn}
\end{figure}

In order to answer these questions, we first characterised the radio afterglow light curves of a synthetic 
population of GRBs constructed so that its $\gamma$-ray properties (flux distributions, energetics/luminosities) 
match those of the real populations of GRBs detected by CGRO/BATSE, \sw\ and \textit{Fermi}. 
GRBs are distributed in redshift between $z = 0$ and 10 according to the GRB formation rate \cite[see][for the different components that are used]{salvaterra12}. 
For each simulated burst, the radio afterglow emission was simulated using a hydrodynamic numerical code (for full details we refer the reader to \citealp{vaneerten11}, and \citealp{vaneerten12}).
The micro-physical parameters (how much energy is distributed to the electrons and magnetic fields, the slope of the electron distribution, and the density of the homogeneous ambient medium) were set in order to reproduce the typical fluxes of the afterglows observed in the optical \citep{melandri14} and radio \citep{ghirlanda13c} bands of a carefully selected sample of bright GRBs observed by 
\sw\ \citep[the so-called BAT6 sample of][]{salvaterra12}. 
These are the brightest GRBs detected by \sw/BAT and they have typical flux densities between 1 and 5 days distributed between 0.1 and 2 mJy at 8.4~GHz \citep{chandra12}. The micro-physical parameters were tuned to reproduce these flux levels (as shown by the red points in the right panel of Fig.~\ref{logn}).
This approach is discussed in detail by \cite{ghirlanda13a}. 
A caveat is the lack of the possible contribution of the reverse shock component, which might 
dominate the emission at early times. Estimates on the detection rates at early 
times should hence be conservative lower limits. Another caveat is that the simulations do not consider the possible supernova contribution, nor the effects of scintillation.

At early times (days after the burst), the emission from the forward shock is suppressed by synchrotron self-absorption, and therefore the flux density due to this component is expected to rise until a \textit{spectral peak} is 
produced by the passage of the frequency (be it the injection frequency or the self-absorption frequency) corresponding to the peak of the spectrum in the telescope observing band. This effectively means that the 
observing strategy should consider the available frequencies and the typical behaviour of the flux of the afterglow 
at those frequencies. 
For instance, the detection rate is maximised if the observations are performed close to the time when the peak of 
the afterglow is reached at a certain frequency. On the other hand, early time observations, which are critical for 
the multi-wavelength study of the GRB emission (coupling radio with optical/NIR, and X-rays), should be planned considering that 
the radio spectrum of the afterglow is still self-absorbed. Therefore, a successful detection is optimised if 
performed at relatively high radio frequencies\footnote{Conversely, the chance of detection of the reverse shock 
component is maximised in the first day.}. 

In Fig.~\ref{logn} (left) we show the cumulative flux distribution of the simulated sample of GRBs. The different 
colours represent different bands of the SKA. The shaded regions represent the cumulative flux distributions if 
observations are performed between 2 and 10 days after the GRB trigger. The peak flux distributions are shown by 
the dashed lines. Two behaviours are clearly visible. 
Firstly, the highest flux is always at the time of the 
second observation (i.e. the emission is still in the self-absorbed region of the spectrum). 
Secondly, the width of the 
region decreases with increasing frequency, because observing in a higher radio band will bring the observation 
closer to the region in the spectrum where most of the flux is emitted, and where the evolution is faster. 
SKA1-MID band 5 (9.2 GHz) is clearly the band that tends to pick a radio afterglow between 2 and 10 days closest to the time of the peak. Another way of looking at the distance of the dashed line from the corresponding shaded area (which again, is maximum for SKA1-LOW, and minimum for the highest SKA1-MID band shown), is a measure of how long to delay an observation in order to maximise the chance of detection in each SKA band.
\begin{table}[!h]
\small
\caption{SKA detection rates of radio afterglows independent from high-energy triggers. For each band we report the 5$\sigma$ sensitivity limit. A standard observation of 12 hours is assumed. $^a$ Refers to ``early science'', an initial implementation with half the sensitivity across all bands. $^b$ The sensitivity is assumed to be 4$\times$ better in the LOW band, and 10$\times$ otherwise.} 
\begin{center}
\begin{tabular}{lccc}
& & SKA1-MID  &\\
\hline \hline
Band &  Freq.& $F_{\rm lim}$  & Rate\\
	 & [GHz]   		  &  [$\mu$Jy]     & [\# sr$^{-1}$ yr$^{-1}$] \\
\hline
LOW			& 0.2			&	3		   	    &	0 	\\	
Band~2	& 1.4			&	1.39		      	    &	300 	\\	
Band~4	& 4.0			&	0.65		      	    &	465 	\\	
Band~5	& 9.2			&	0.75		      	    &	465 	\\	
\hline 
\end{tabular}
\begin{tabular}{cc}
SKA1-MID E.S.$^a$ & SKA2$^b$\\
\hline \hline
Rate & Rate \\
~[\# sr$^{-1}$ yr$^{-1}$] & [\# sr$^{-1}$ yr$^{-1}$]  \\
\hline
0 & 2\\
170	& 465\\
450	& 465\\
465	& 465\\
\hline 
\end{tabular}\label{tab:superswift}
\end{center}
\end{table}

Alternatively, one can visualise this behaviour by investigating the flux density at which a GRB afterglow peaks 
when the transition from optically thick-to-thin happens. Fig.~\ref{logn} (right) shows a plot of the peak fluxes 
versus the time at which the frequency corresponding to the peak of the spectrum crosses the band (different colours represent the simulated SKA bands). 
The more energetic GRBs peak at later times, with the lower frequencies lagging the high ones. The brighter bursts 
populate the upper right part of the distributions, and can indeed be brighter than 0.1 mJy at 1.4~GHz even months 
after the prompt emission has disappeared. However they are very rare, representing less than 1\% of the whole population. 
We note that this is probably a lower limit on the actual fraction of these highly luminous afterglows, given that 
small number statistics dominate the tail of the simulated distributions. 
We have calculated the rates of GRB afterglows that we expect to be able to observe at the time of their peak, depending on the level of sensitivity\footnote{ Sensitivities are calculated from the SKA1 system baseline design documents, see \cite{dewdney13}.} the SKA1 and SKA2 will reach in each band. Table~\ref{tab:superswift} summarises these results for the whole synthetic population of afterglows (i.e. regardless of their $\gamma$-ray properties). Given that the logN-logF is almost flat at sub-$\mu$Jy levels, in particular for the high frequency bands, there is no big improvement in expected rates from SKA1 to SKA2. The catch is that in order to detect a GRB pointed to Earth, we will always need a trigger. On the other hand, for the GRBs that we will be able to detect, the SKA will be able to sample to lightcurve down to the $\mu$Jy level, producing finely sampled lightcurves, that current facilities are able to produce just for the very close-by GRBs.

In the past, the observation strategy for GRB afterglows was driven more by the goal of having coeval multi-wavelength observations across the spectrum. The need of being on target as soon as possible in the optical, NIR, and X-rays possibly drove the attempt of having a detection in the very first few days. A small number of bursts were detected and followed-up to extremely late times \citep[see][]{frail97, frail04, berger04, vanderhorst05, vanderhorst08}, but the large majority ($>70\%$) of radio observations resulted in non-detections overall. Note that when just the ($\gamma$-ray) brighter population of bursts was considered, the ratio of positive detections rose to $\sim 50\%$ \citep{ghirlanda13c}, which hints that the SKA will indeed be able to probe the hidden population of radio afterglows (Anderson et al., 2014 --in preparation-- will present the first systematic radio follow-up of \sw\ GRBs at GHz frequencies).
	
		\subsection{GRB calorimetry: accessing the true energy budget} \label{sect:nonrel}

One of the key characteristics of the GRB phenomenon that has eluded direct measurements is the estimate of the true energy budget of the central engine that powers the explosion. 
The isotropic equivalent energy emitted in $\gamma$-rays (\eiso) is just a proxy of the true GRB energy.
Indeed, due to the collimated structure of GRBs, the energy $E_{\gamma}$ emitted just in $\gamma$-rays is a factor $\approx~$\th$^2/2$ smaller than that derived assuming isotropy, where \th\ is the jet angle. Moreover, the measure of the jet angle can be marred by observational biases and possibly by additional components at play in the external shock emission. Additionally, the energy emitted in the high-energy band is a fraction of the kinetic energy (\ekin), initially available.

According to the standard model, the fireball becomes non-relativistic (NR) at a time $t_{\rm NR}$, which is long after the prompt emission has faded away.
The theoretical prediction of when this transition should occur is subject to debate, and consequently there is roughly an order of magnitude in the range of values that $t_{\rm NR}$ can take, from hundreds \citep[e.g.][]{livio00, sironi13} to thousands of days \cite[e.g.][]{piran04, zhang09, wygoda11, vaneerten12}. For this Chapter, we have adopted the ``best case'' scenario, with the caveat that this is subject to further improvements.
Following \cite{livio00}, the transition happens at 
$t_{\rm NR} \sim 275(1+z)(n/1~\rm{cm}^{-3})^{-1/3}$(\ekin$/10^{53}~\rm{erg})^{1/3}$\,days.  
The advantage of performing energetic estimates (calorimetry) at such late times is that the afterglow emission becomes independent from the bulk Lorentz factor (which becomes $\approx1$). 
This allows us to derive the true GRB kinetic energy \citep[e.g.][]{frail2000, shivvers11}, which is directly related to the energy emitted in $\gamma$-rays during the prompt phase through a radiative efficiency factor $\eta$. This is related to the measurable \eiso\ by \ekinj$\approx [$\th$^2/(2\eta)]\times$\eiso. 
This is where the power of these observations becomes evident: assuming the efficiency $\eta$ of the shocks will give us an independent measure of the jet opening angle; conversely for the cases in which there will be a measure of the jet opening angle (e.g. from the achromatic break in the X-ray and optical afterglow), we will be able to compute the shock efficiency.
\begin{figure}[!ht]
\begin{center}
\includegraphics[width=8cm,trim=40 10 20 10]{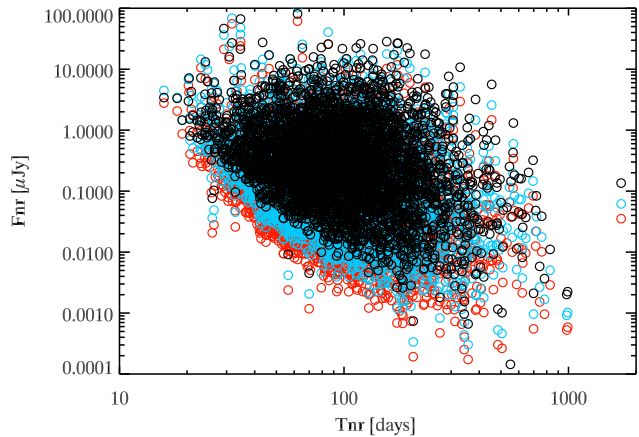}
\caption{Transition to the NR regime: the flux is calculated at time of the crossing to the NR regime, calculated as per Livio \& Waxmann, 2001. The colour code is as per Fig.~3.}
\label{fnr_tnr}
\end{center}
\end{figure}
 
In addition, observations before and after $t_{\rm NR}$ are probes of the ambient medium profile. In the simple case of a
 homogeneous density ambient medium, a peculiar flattening of the lightcurve is predicted by the standard model. 
In the simple case of uniform density ISM, a flattening of $\Delta \alpha = (21-5p)/10$ is expected for the lightcurve across this transition \citep{frail05}, for the standard values of the electron distribution slope $p$. 
Interestingly, a steepening could also potentially be observed at times after $t_{\rm NR}$, and that would also help us constrain the microphysical parameter $p$.  
In the case of a wind-like medium, the transition is expected at later times, with a distinct slope variation of the lightcurve. Therefore observing across the NR transition is a powerful diagnostic of the circumburst medium density profile.

We have computed the expected flux density at the NR transition for the SKA bands 2, 4, and 5, see Fig.~\ref{fnr_tnr}. 
Consistently with the observations available so far, only a handful of objects reach the sensitivity limit of current facilities (a few $\mu$Jy at best). We foresee that SKA1-MID will be able to observe a minor fraction of these (of the order of 2--5\%, down to 0.9--2\% for the ``early science'' phase), but 
that the full SKA will routinely observe a significant fraction (15--25\%) of the whole GRB afterglow population at the NR transition.Even for the full SKA though, a substantial (75--85\%) fraction of sources will be undetectable. We calculated the fractions based on the number of afterglows above a 12-hour-long observation (5$\sigma$), over the total population. 

		\subsection{The case for a very sensitive high--energy telescope}

\begin{figure}[!ht]
\begin{center}
\includegraphics[width=8cm,trim=40 10 20 10]{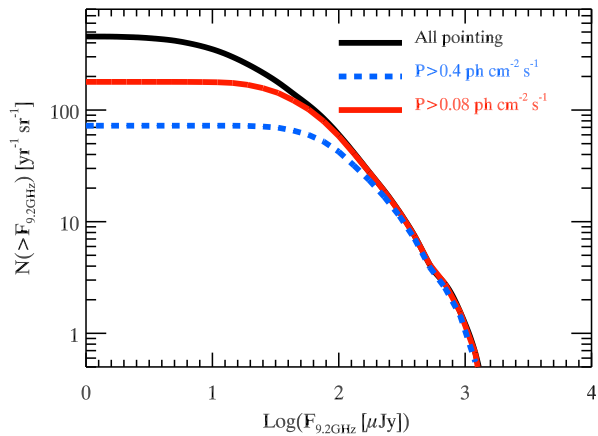}
\caption{
LogN--logF of SKA1-MID band 5 (9.2 GHz) showing the flux at the peak of the light curve for the population of simulated GRBs. The red solid and blue dashed lines represent the population of GRBs that have a $\gamma$-ray photon peak flux (P) greater than 0.08 and 0.4 ph cm$^{-2}$ s$^{-1}$, respectively. The latter represents the synthetic population of bursts that can trigger \sw.~The improvement of radio flux sensitivity assured by the SKA can be exploited only if a $\gamma$-ray detector more sensitive (in the case shown, we assume a five-fold improvement in trigger sensitivity) than \sw\ will be operating in the SKA era. The distance between the red and blue line shows how many GRB afterglows, whose radio flux density is accessible by the SKA, we will miss out unless there is a GRB trigger. The black line shows all GRBs pointing to the Earth, irrespectively of the their $\gamma$-ray flux.}
\label{superswift}
\end{center}
\end{figure}

The sensitivity of the SKA will allow us to observe almost the complete population of GRBs, provided that a $\gamma$-ray instrument,  more sensitive than \sw\ will be operational in the SKA era, and provided that a GRB localisation to a few arcsec spatial resolution will be available.
We have explored the case of a burst monitor with a 
 ($\sim5\times$) greater sensitivity than that of the \sw/BAT. As can be seen in Fig.~\ref{superswift}, the bursts that can be seen in band~5 (9.2 GHz) of SKA1-MID are a larger fraction of the whole population with respect to the ones that will trigger \sw\, provided that a $\gamma$-ray trigger is supplied. 

In short, the future absence of a burst monitor with an improved sensitivity than that of the \sw/BAT, would imply that the rate of GRB afterglows 
that are observable at the $\mu$-Jy level is comparable to the the ones we can already observe today (i.e. $\sim 50$ sr$^{-1}$ yr$^{-1}$). In conclusion, while trigger-dependent detection rates might not dramatically improve in the SKA era, the sampling of the lightcurves of the GRBs that will produce a $\gamma$-ray trigger will be far better than that available today.

	\section{Revealing the orphan afterglow population}\label{sect:orphan}
Most GRBs do not have their jets pointing towards us. As a result they are undetectable in the $\gamma$-ray band but
should be detectable as `orphan afterglow (OA)'. Given the typical jet opening angles of GRBs, the population
of OAs should outnumber the pointing GRB population by two orders of magnitude. 

Jet opening angles have been derived by the afterglow light curve break \citep{frail01, ghirlanda04a} and have a distribution clustered at \th$\sim5$ deg. It is expected that, purely by geometric arguments, for each GRB pointing towards the Earth that can potentially trigger a $\gamma$-ray instrument, there are $\sim2$(\th$/0.1)^{-2}\approx 260$ events directed in all other directions in the sky. More realistically, \cite{ghirlanda14} have estimated that --- given that \th\ has a distribution of values --- the ratio of OAs to GRBs pointing to us is $\sim 40$. 
These bursts are undetectable in the high energy band because are viewed at angles \thv~$>$~\th, but are expected to be potentially observable at optical and radio wavelengths from the time when, during the afterglow evolution, their Lorentz factors slows down to $\Gamma \sim$\thv$^{-1}$.
These events should therefore lack their prompt emission (and hence have been dubbed `orphan afterglows') and should show up in deep optical and radio transient surveys. The characterisation of their properties, the feature of their lightcurves, the typical timescales and peak fluxes are of particular interest for the deep, large surveys of the future facilities like SKA1-SUR, since this class of transients could (i) be difficult to disentangle from other classes of transients, and (ii) constitute a considerable fraction of transient events in the sky. 
Up until today the quest for their discovery has resulted in upper \citep[\th\lsim22$^{\circ}$][]{levinson02} or 
lower \citep[\th\gsim0.8$^{\circ}$][]{soderberg06} limits on the GRB opening angles. Some peculiar transients in past surveys could have been genuine OAs both in the optical \citep{rau06, zou07, malacrino07} and radio \citep[for a recent reveiw on blind surveys see][]{murphy13} regime, but none have been confirmed so far. 

\begin{figure}[!ht]
\begin{center}
\includegraphics[width=8cm,trim=40 10 20 10]{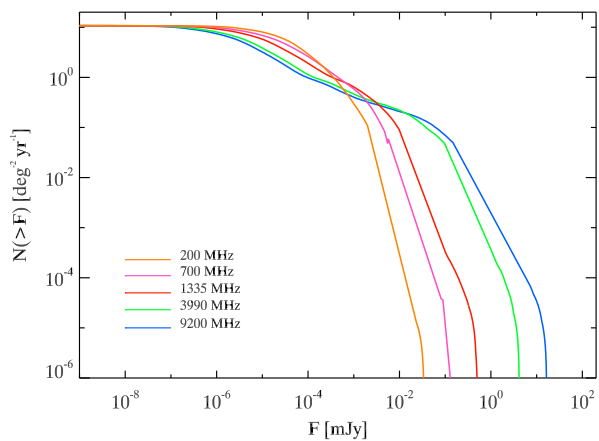}
\caption{
Cumulative flux distribution of orphan afterglows at the frequencies of SKA1-MID/SKA1-SUR and SKA1-LOW are shown. To compute the number of expected OA detections, as detailed in the text, we assumed a deep (1000~hour), all-sky (3$\pi$) survey. The rates can be rescaled easily since they are linearly proportional to the sky fraction covered.
}
\label{logn_oa}
\end{center}
\end{figure}

The evolution of the OA lightcurve differs from that of a normal afterglow before reaching its peak flux. In fact, an OA should start to be potentially observable\footnote{Under the simplified assumption of a sharp edged cone shape for the jet.} when the edge of the jet  intercepts the line of sight of the observer: this happens when the fireball has decelerated to $\Gamma = 1/\sin($\thv$-$\th). 
In this respect, the OA is driven by geometry because it is a \textit{dynamical peak}, not a spectral peak generated by the passage of the peak of the spectrum through the observing band. This happens at $\Gamma = 1/\sin($\thv), and the afterglow evolves like that of a standard GRB thereafter. 
\cite{ghirlanda14} have estimated that the distribution of peak times is very broad and peaks at $\sim 1000$\,days. 
In addition, we show in Fig.~\ref{logn_oa} the cumulative flux distribution of a simulated sample of OAs in the SKA1 bands. The rates are expressed in units of deg$^{-2}$\,yr$^{-1}$, so that the conversion to the prediction for different fields of view will be straightforward.
The high (low) flux tail of the logN-logF are consistent with a slope of $-1.7$ ($-0.4$) roughly in all bands.
The GRBs with small opening angles dominate the logN-logF in the high flux tail; conversely the low flux tail is due to the cumulative contribution of GRBs with progressively large opening angles. 
We calculated from Fig.~\ref{logn_oa} how many OAs an all-sky survey with SKA1-SUR should be able to detect each week in every PAF band. This result is summarised in Table~\ref{tab:orphan}.

\begin{table}[!ht]
\small \centering
\caption{SKA1-SUR detection rates of orphan afterglows in the SKA1-SUR bands. 
The rates are derived from the flux density distributions shown in Fig.~6, assuming 1000~hours of integration time, all-sky ($3\pi$), 5$\sigma$ and given in units of $3\pi^{-1}$ week$^{-1}$. 
$F_{\rm lim}$ is in units of $\mu$Jy.
$^{a}$~Refers to ``early science'', an initial implementation with half the sensitivity.
$^b$~The sensitivity is assumed to be 10~$\times$ better for SKA2.}
\begin{tabular}{cccccccc}
& & SKA1-SUR && SKA1-SUR E.S.$^a$ && SKA2$^b$ &\\
\hline \hline
Band &  Freq. [GHz] & 	$F_{\rm lim}$ & Rate &$F_{\rm lim}$ & Rate& $F_{\rm lim}$ & Rate\\
\hline
PAF band~1	& 0.7 	&	1.54 &	357	&	3.08&	178	&	0.2&	1190	\\	
PAF band~2	& 1.4	&	0.92 &	387	&	1.84&	297	&	0.9&	1011	\\		
PAF band~3	& 4.0	&	1.23 &	297	&	2.46&	237	&	0.6&	892	\\	
\hline 
\end{tabular}\label{tab:orphan}
\end{table}

The possibility of revealing the population of orphan afterglows by SKA through its survey is tantalizing. Our estimates suggests that at radio frequencies orphan afterglows should be relatively long-term transients with typical duration above the expected flux limit of the SKA survey of hundreds of days. The challenge will be to distinguish the population of orphans from other possible sources (non-GRBs) that produce such slow transients. The unique possibility offered by the SKA multifrequency observations will be to study the radio spectrum of these transients. Moreover, as discussed more in detail in the Chapter on the VLBI capabilities of the SKA (Paragi et al., ``Very Long Baseline Interferometry with the SKA'', in proceedings of ``Advancing Astrophysics with the Square Kilometre Array'', PoS(AASKA14)143), the longest baselines will secure sub-mas resolution. This is turn will mean that, at least for the OAs with sufficient signal-to-noise we will be able to measure the expansion velocity of the ejecta. This will be an additional tool to distinguish transients powered by a central engine from radio SNe.


	\section{Conclusion and the road to the SKA}
The SKA era will be transformational for the study of GRBs, because it will allow us to
observe close to 100\% of the detected GRBs directed towards the Earth. 
In addition to very high detection \emph{rates}, if compared to the state of the art, it will be 
able to produce well-sampled radio lightcurves, bringing radio-astronomy to the same level of 
completeness obtained in the last decade at higher (optical, X-ray) frequencies.

For a quarter of all GRBs with a jet pointing to the Earth, the SKA will be able to trace the 
lightcurve up to very late times, even across the non-relativistic transition. We will get an 
unprecedented insight into the true energy budget of GRB engines by means of these very late-time 
observations. Additionally, these observations will probe both the macrophysics of the ambient medium 
(e.g. the density profile) and the microphysics of the shocks, which are believed to be at the 
basis of afterglow emission.

The ability to both increase the sheer rate of detections, and to follow the afterglow lightcurve to very late times will naturally depend on 
the availability of a $\gamma$-ray trigger with arcsecond resolution.

Finally we will be able to observe the so far elusive class of ``orphan afterglows'', which are 
predicted to exceed the standard afterglow rate by orders of magnitude.
The full SKA telescope should detect up to $\gsim 1000$ of these slowly evolving transients 
\emph{every week}. 

On the way to the SKA1, the ``early science'' phase will roughly triple the rate of observed afterglows from GRBs pointed towards Earth. The next steps forward will be taken by surveys on SKA pathfinder instruments
such as the Australian SKA Pathfinder \citep[ASKAP;][]{johnston07}. The VAST survey \citep{murphy13} at GHz frequencies will be first of its kind to reach a uniform sub-mJy sensitivity in the whole southern hemisphere.
VAST will not have the sensitivity to target the bulk of the 
afterglow population, but it will be the benchmark for the SKA era blind surveys of slow transients. 

\acknowledgments 

TM and DB acknowledge the support of the Australian Research Council (ARC) through
Discovery Project DP110102034, and, with BMG, through the ARC Centre
of Excellence for All-sky Astrophysics (CE110001020). 
GG acknowledges INAF PRIN 2011.
AJvdH and RAMJW acknowledge support from the European Research Council via Advanced Investigator Grant no. 247295

\setlength{\bibsep}{0.0pt}
\bibliographystyle{apj}

\begin{thebibliography}{97}
\expandafter\ifx\csname natexlab\endcsname\relax\def\natexlab#1{#1}\fi

\bibitem[{{Ackermann} {et~al.}(2013){Ackermann}, {Ajello}, {Asano}, {Axelsson},
  {Baldini}, {Ballet}, {Barbiellini}, {Bastieri}, {Bechtol}, {Bellazzini},
  {Bhat}, {Bissaldi}, {Bloom}, {Bonamente}, {Bonnell}, {Bouvier}, {Brandt},
  {Bregeon}, {Brigida}, {Bruel}, {Buehler}, {Burgess}, {Buson}, {Byrne},
  {Caliandro}, {Cameron}, {Caraveo}, {Cecchi}, {Charles}, {Chaves},
  {Chekhtman}, {Chiang}, {Chiaro}, {Ciprini}, {Claus}, {Cohen-Tanugi},
  {Connaughton}, {Conrad}, {Cutini}, {D'Ammando}, {de Angelis}, {de Palma},
  {Dermer}, {Desiante}, {Digel}, {Dingus}, {Di Venere}, {Drell},
  {Drlica-Wagner}, {Dubois}, {Favuzzi}, {Ferrara}, {Fitzpatrick}, {Foley},
  {Franckowiak}, {Fukazawa}, {Fusco}, {Gargano}, {Gasparrini}, {Gehrels},
  {Germani}, {Giglietto}, {Giommi}, {Giordano}, {Giroletti}, {Glanzman},
  {Godfrey}, {Goldstein}, {Granot}, {Grenier}, {Grove}, {Gruber}, {Guiriec},
  {Hadasch}, {Hanabata}, {Hayashida}, {Horan}, {Hou}, {Hughes}, {Inoue},
  {Jackson}, {Jogler}, {J{\'o}hannesson}, {Johnson}, {Johnson}, {Kamae},
  {Kataoka}, {Kawano}, {Kippen}, {Kn{\"o}dlseder}, {Kocevski}, {Kouveliotou},
  {Kuss}, {Lande}, {Larsson}, {Latronico}, {Lee}, {Longo}, {Loparco},
  {Lovellette}, {Lubrano}, {Massaro}, {Mayer}, {Mazziotta}, {McBreen},
  {McEnery}, {McGlynn}, {Michelson}, {Mizuno}, {Moiseev}, {Monte}, {Monzani},
  {Moretti}, {Morselli}, {Murgia}, {Nemmen}, {Nuss}, {Nymark}, {Ohno},
  {Ohsugi}, {Omodei}, {Orienti}, {Orlando}, {Paciesas}, {Paneque}, {Panetta},
  {Pelassa}, {Perkins}, {Pesce-Rollins}, {Piron}, {Pivato}, {Porter}, {Preece},
  {Racusin}, {Rain{\`o}}, {Rando}, {Rau}, {Razzano}, {Razzaque}, {Reimer},
  {Reimer}, {Reposeur}, {Ritz}, {Romoli}, {Roth}, {Ryde}, {Saz Parkinson},
  {Schalk}, {Sgr{\`o}}, {Siskind}, {Sonbas}, {Spandre}, {Spinelli}, {Suson},
  {Tajima}, {Takahashi}, {Takeuchi}, {Tanaka}, {Thayer}, {Thayer}, {Thompson},
  {Tibaldo}, {Tierney}, {Tinivella}, {Torres}, {Tosti}, {Troja}, {Tronconi},
  {Usher}, {Vandenbroucke}, {van der Horst}, {Vasileiou}, {Vianello}, {Vitale},
  {von Kienlin}, {Winer}, {Wood}, {Wood}, {Xiong}, \& {Yang}}]{ackermann2013}
{Ackermann}, M., {et~al.} 2013, \apjs, 209, 11

\bibitem[{{Ackermann} {et~al.}(2014){Ackermann}, {Ajello}, {Asano}, {Atwood},
  {Axelsson}, {Baldini}, {Ballet}, {Barbiellini}, {Baring}, {Bastieri},
  {Bechtol}, {Bellazzini}, {Bissaldi}, {Bonamente}, {Bregeon}, {Brigida},
  {Bruel}, {Buehler}, {Burgess}, {Buson}, {Caliandro}, {Cameron}, {Caraveo},
  {Cecchi}, {Chaplin}, {Charles}, {Chekhtman}, {Cheung}, {Chiang}, {Chiaro},
  {Ciprini}, {Claus}, {Cleveland}, {Cohen-Tanugi}, {Collazzi}, {Cominsky},
  {Connaughton}, {Conrad}, {Cutini}, {D'Ammando}, {de Angelis}, {DeKlotz}, {de
  Palma}, {Dermer}, {Desiante}, {Diekmann}, {Di Venere}, {Drell},
  {Drlica-Wagner}, {Favuzzi}, {Fegan}, {Ferrara}, {Finke}, {Fitzpatrick},
  {Focke}, {Franckowiak}, {Fukazawa}, {Funk}, {Fusco}, {Gargano}, {Gehrels},
  {Germani}, {Gibby}, {Giglietto}, {Giles}, {Giordano}, {Giroletti}, {Godfrey},
  {Granot}, {Grenier}, {Grove}, {Gruber}, {Guiriec}, {Hadasch}, {Hanabata},
  {Harding}, {Hayashida}, {Hays}, {Horan}, {Hughes}, {Inoue}, {Jogler},
  {J{\'o}hannesson}, {Johnson}, {Kawano}, {Kn{\"o}dlseder}, {Kocevski}, {Kuss},
  {Lande}, {Larsson}, {Latronico}, {Longo}, {Loparco}, {Lovellette}, {Lubrano},
  {Mayer}, {Mazziotta}, {McEnery}, {Michelson}, {Mizuno}, {Moiseev}, {Monzani},
  {Moretti}, {Morselli}, {Moskalenko}, {Murgia}, {Nemmen}, {Nuss}, {Ohno},
  {Ohsugi}, {Okumura}, {Omodei}, {Orienti}, {Paneque}, {Pelassa}, {Perkins},
  {Pesce-Rollins}, {Petrosian}, {Piron}, {Pivato}, {Porter}, {Racusin},
  {Rain{\`o}}, {Rando}, {Razzano}, {Razzaque}, {Reimer}, {Reimer}, {Ritz},
  {Roth}, {Ryde}, {Sartori}, {Parkinson}, {Scargle}, {Schulz}, {Sgr{\`o}},
  {Siskind}, {Sonbas}, {Spandre}, {Spinelli}, {Tajima}, {Takahashi}, {Thayer},
  {Thayer}, {Thompson}, {Tibaldo}, {Tinivella}, {Torres}, {Tosti}, {Troja},
  {Usher}, {Vandenbroucke}, {Vasileiou}, {Vianello}, {Vitale}, {Winer}, {Wood},
  {Yamazaki}, {Younes}, {Yu}, {Zhu}, {Bhat}, {Briggs}, {Byrne}, {Foley},
  {Goldstein}, {Jenke}, {Kippen}, {Kouveliotou}, {McBreen}, {Meegan},
  {Paciesas}, {Preece}, {Rau}, {Tierney}, {van der Horst}, {von Kienlin},
  {Wilson-Hodge}, {Xiong}, {Cusumano}, {La Parola}, \&
  {Cummings}}]{ackermann2014}
---. 2014, Science, 343, 42

\bibitem[{{Akerlof} {et~al.}(1999){Akerlof}, {Balsano}, {Barthelmy}, {Bloch},
  {Butterworth}, {Casperson}, {Cline}, {Fletcher}, {Frontera}, {Gisler},
  {Heise}, {Hills}, {Kehoe}, {Lee}, {Marshall}, {McKay}, {Miller}, {Piro},
  {Priedhorsky}, {Szymanski}, \& {Wren}}]{akerlof1999}
{Akerlof}, C., {et~al.} 1999, \nat, 398, 400

\bibitem[{{Anderson} {et~al.}(2014){Anderson}, {van der Horst}, {Staley},
  {Fender}, {Wijers}, {Scaife}, {Rumsey}, {Titterington}, {Rowlinson}, \&
  {Saunders}}]{anderson14}
{Anderson}, G.~E., {et~al.} 2014, \mnras, 440, 2059

\bibitem[{{Bannister} {et~al.}(2012){Bannister}, {Murphy}, {Gaensler}, \&
  {Reynolds}}]{bannister2012}
{Bannister}, K.~W., {Murphy}, T., {Gaensler}, B.~M., \& {Reynolds}, J.~E. 2012,
  \apj, 757, 38

\bibitem[{{Berger} {et~al.}(2004){Berger}, {Kulkarni}, \& {Frail}}]{berger04}
{Berger}, E., {Kulkarni}, S.~R., \& {Frail}, D.~A. 2004, \apj, 612, 966

\bibitem[{{Berger} {et~al.}(2003){Berger}, {Kulkarni}, {Frail}, \&
  {Soderberg}}]{berger2003}
{Berger}, E., {Kulkarni}, S.~R., {Frail}, D.~A., \& {Soderberg}, A.~M. 2003,
  \apj, 599, 408

\bibitem[{{Bietenholz} {et~al.}(2014){Bietenholz}, {De Colle}, {Granot},
  {Bartel}, \& {Soderberg}}]{bietenholz2014}
{Bietenholz}, M.~F., {De Colle}, F., {Granot}, J., {Bartel}, N., \&
  {Soderberg}, A.~M. 2014, \mnras, 440, 821

\bibitem[{{Cenko} {et~al.}(2010){Cenko}, {Frail}, {Harrison}, {Kulkarni},
  {Nakar}, {Chandra}, {Butler}, {Fox}, {Gal-Yam}, {Kasliwal}, {Kelemen},
  {Moon}, {Ofek}, {Price}, {Rau}, {Soderberg}, {Teplitz}, {Werner}, {Bock},
  {Bloom}, {Starr}, {Filippenko}, {Chevalier}, {Gehrels}, {Nousek}, \&
  {Piran}}]{cenko2010}
{Cenko}, S.~B., {et~al.} 2010, \apj, 711, 641

\bibitem[{{Cenko} {et~al.}(2011){Cenko}, {Frail}, {Harrison}, {Haislip},
  {Reichart}, {Butler}, {Cobb}, {Cucchiara}, {Berger}, {Bloom}, {Chandra},
  {Fox}, {Perley}, {Prochaska}, {Filippenko}, {Glazebrook}, {Ivarsen},
  {Kasliwal}, {Kulkarni}, {LaCluyze}, {Lopez}, {Morgan}, {Pettini}, \&
  {Rana}}]{cenko2011}
---. 2011, \apj, 732, 29

\bibitem[{{Cenko} {et~al.}(2013){Cenko}, {Kulkarni}, {Horesh}, {Corsi}, {Fox},
  {Carpenter}, {Frail}, {Nugent}, {Perley}, {Gruber}, {Gal-Yam}, {Groot},
  {Hallinan}, {Ofek}, {Rau}, {MacLeod}, {Miller}, {Bloom}, {Filippenko},
  {Kasliwal}, {Law}, {Morgan}, {Polishook}, {Poznanski}, {Quimby}, {Sesar},
  {Shen}, {Silverman}, \& {Sternberg}}]{cenko13}
---. 2013, \apj, 769, 130

\bibitem[{{Chandra} \& {Frail}(2012)}]{chandra12}
{Chandra}, P., \& {Frail}, D.~A. 2012, \apj, 746, 156

\bibitem[{{Chandra} {et~al.}(2008){Chandra}, {Cenko}, {Frail}, {Chevalier},
  {Macquart}, {Kulkarni}, {Bock}, {Bertoldi}, {Kasliwal}, {Fox}, {Price},
  {Berger}, {Soderberg}, {Harrison}, {Gal-Yam}, {Ofek}, {Rau}, {Schmidt},
  {Cameron}, {Cowie}, {Cowie}, {Roth}, {Dopita}, {Peterson}, \&
  {Penprase}}]{chandra2008}
{Chandra}, P., {et~al.} 2008, \apj, 683, 924

\bibitem[{{Costa} {et~al.}(1997){Costa}, {Frontera}, {Heise}, {Feroci}, {in't
  Zand}, {Fiore}, {Cinti}, {Dal Fiume}, {Nicastro}, {Orlandini}, {Palazzi},
  {Rapisarda\#}, {Zavattini}, {Jager}, {Parmar}, {Owens}, {Molendi},
  {Cusumano}, {Maccarone}, {Giarrusso}, {Coletta}, {Antonelli}, {Giommi},
  {Muller}, {Piro}, \& {Butler}}]{costa1997}
{Costa}, E., {et~al.} 1997, \nat, 387, 783

\bibitem[{{Covino} {et~al.}(1999){Covino}, {Lazzati}, {Ghisellini}, {Saracco},
  {Campana}, {Chincarini}, {di Serego}, {Cimatti}, {Vanzi}, {Pasquini},
  {Haardt}, {Israel}, {Stella}, \& {Vietri}}]{covino1999}
{Covino}, S., {et~al.} 1999, \aap, 348, L1

\bibitem[{{Cucchiara} {et~al.}(2011){Cucchiara}, {Levan}, {Fox}, {Tanvir},
  {Ukwatta}, {Berger}, {Kr{\"u}hler}, {K{\"u}pc{\"u} Yolda{\c s}}, {Wu},
  {Toma}, {Greiner}, {Olivares}, {Rowlinson}, {Amati}, {Sakamoto}, {Roth},
  {Stephens}, {Fritz}, {Fynbo}, {Hjorth}, {Malesani}, {Jakobsson}, {Wiersema},
  {O'Brien}, {Soderberg}, {Foley}, {Fruchter}, {Rhoads}, {Rutledge}, {Schmidt},
  {Dopita}, {Podsiadlowski}, {Willingale}, {Wolf}, {Kulkarni}, \&
  {D'Avanzo}}]{cucchiara2011}
{Cucchiara}, A., {et~al.} 2011, \apj, 736, 7

\bibitem[{{De Colle} {et~al.}(2012){De Colle}, {Ramirez-Ruiz}, {Granot}, \&
  {Lopez-Camara}}]{decolle2012}
{De Colle}, F., {Ramirez-Ruiz}, E., {Granot}, J., \& {Lopez-Camara}, D. 2012,
  \apj, 751, 57

\bibitem[{{Dewdney} {et~al.}(2013){Dewdney}, {Turner}, {Millenaar}, {McCool},
  {Lazio}, \& {Cornwell}}]{dewdney13}
{Dewdney}, P., {Turner}, W., {Millenaar}, R., {McCool}, R., {Lazio}, J., \&
  {Cornwell}, T. 2013, SKA Documents, Document number SKA-TEL-SK0-DD-001,
  Revision 1

\bibitem[{{Eichler} {et~al.}(1989){Eichler}, {Livio}, {Piran}, \&
  {Schramm}}]{eichler1989}
{Eichler}, D., {Livio}, M., {Piran}, T., \& {Schramm}, D.~N. 1989, \nat, 340,
  126

\bibitem[{{Frail} {et~al.}(1997){Frail}, {Kulkarni}, {Nicastro}, {Feroci}, \&
  {Taylor}}]{frail97}
{Frail}, D.~A., {Kulkarni}, S.~R., {Nicastro}, L., {Feroci}, M., \& {Taylor},
  G.~B. 1997, \nat, 389, 261

\bibitem[{{Frail} {et~al.}(2004){Frail}, {Metzger}, {Berger}, {Kulkarni}, \&
  {Yost}}]{frail04}
{Frail}, D.~A., {Metzger}, B.~D., {Berger}, E., {Kulkarni}, S.~R., \& {Yost},
  S.~A. 2004, \apj, 600, 828

\bibitem[{{Frail} {et~al.}(2005){Frail}, {Soderberg}, {Kulkarni}, {Berger},
  {Yost}, {Fox}, \& {Harrison}}]{frail05}
{Frail}, D.~A., {Soderberg}, A.~M., {Kulkarni}, S.~R., {Berger}, E., {Yost},
  S., {Fox}, D.~W., \& {Harrison}, F.~A. 2005, \apj, 619, 994

\bibitem[{{Frail} {et~al.}(2000){Frail}, {Waxman}, \& {Kulkarni}}]{frail2000}
{Frail}, D.~A., {Waxman}, E., \& {Kulkarni}, S.~R. 2000, \apj, 537, 191

\bibitem[{{Frail} {et~al.}(2001){Frail}, {Kulkarni}, {Sari}, {Djorgovski},
  {Bloom}, {Galama}, {Reichart}, {Berger}, {Harrison}, {Price}, {Yost},
  {Diercks}, {Goodrich}, \& {Chaffee}}]{frail01}
{Frail}, D.~A., {et~al.} 2001, \apjl, 562, L55

\bibitem[{{Gehrels} {et~al.}(2005){Gehrels}, {Sarazin}, {O'Brien}, {Zhang},
  {Barbier}, {Barthelmy}, {Blustin}, {Burrows}, {Cannizzo}, {Cummings}, {Goad},
  {Holland}, {Hurkett}, {Kennea}, {Levan}, {Markwardt}, {Mason}, {Meszaros},
  {Page}, {Palmer}, {Rol}, {Sakamoto}, {Willingale}, {Angelini}, {Beardmore},
  {Boyd}, {Breeveld}, {Campana}, {Chester}, {Chincarini}, {Cominsky},
  {Cusumano}, {de Pasquale}, {Fenimore}, {Giommi}, {Gronwall}, {Grupe}, {Hill},
  {Hinshaw}, {Hjorth}, {Hullinger}, {Hurley}, {Klose}, {Kobayashi},
  {Kouveliotou}, {Krimm}, {Mangano}, {Marshall}, {McGowan}, {Moretti},
  {Mushotzky}, {Nakazawa}, {Norris}, {Nousek}, {Osborne}, {Page}, {Parsons},
  {Patel}, {Perri}, {Poole}, {Romano}, {Roming}, {Rosen}, {Sato}, {Schady},
  {Smale}, {Sollerman}, {Starling}, {Still}, {Suzuki}, {Tagliaferri},
  {Takahashi}, {Tashiro}, {Tueller}, {Wells}, {White}, \&
  {Wijers}}]{gehrels2005}
{Gehrels}, N., {et~al.} 2005, \nat, 437, 851

\bibitem[{{Ghirlanda} {et~al.}(2004){Ghirlanda}, {Ghisellini}, \&
  {Lazzati}}]{ghirlanda04a}
{Ghirlanda}, G., {Ghisellini}, G., \& {Lazzati}, D. 2004, \apj, 616, 331

\bibitem[{{Ghirlanda} {et~al.}(2013{\natexlab{a}}){Ghirlanda}, {Salvaterra},
  {Burlon}, {Campana}, {Melandri}, {Bernardini}, {Covino}, {D'Avanzo},
  {D'Elia}, {Ghisellini}, {Nava}, {Prandoni}, {Sironi}, {Tagliaferri},
  {Vergani}, \& {Wolter}}]{ghirlanda13c}
{Ghirlanda}, G., {et~al.} 2013{\natexlab{a}}, \mnras, 435, 2543

\bibitem[{{Ghirlanda} {et~al.}(2013{\natexlab{b}}){Ghirlanda}, {Ghisellini},
  {Salvaterra}, {Nava}, {Burlon}, {Tagliaferri}, {Campana}, {D'Avanzo}, \&
  {Melandri}}]{ghirlanda13a}
---. 2013{\natexlab{b}}, \mnras, 428, 1410

\bibitem[{{Ghirlanda} {et~al.}(2014){Ghirlanda}, {Burlon}, {Ghisellini},
  {Salvaterra}, {Bernardini}, {Campana}, {Covino}, {D'Avanzo}, {D'Elia},
  {Melandri}, {Murphy}, {Nava}, {Vergani}, \& {Tagliaferri}}]{ghirlanda14}
---. 2014, \pasa, 31, 22

\bibitem[{{Goodman}(1997)}]{goodman1997}
{Goodman}, J. 1997, \nat, 2, 449

\bibitem[{{Granot} \& {Loeb}(2003)}]{granot2003}
{Granot}, J., \& {Loeb}, A. 2003, \apjl, 593, L81

\bibitem[{{Granot} {et~al.}(2005){Granot}, {Ramirez-Ruiz}, \&
  {Loeb}}]{granot2005}
{Granot}, J., {Ramirez-Ruiz}, E., \& {Loeb}, A. 2005, \apj, 618, 413

\bibitem[{{Granot} \& {van der Horst}(2014)}]{granot2014}
{Granot}, J., \& {van der Horst}, A.~J. 2014, \pasa, 31, 8

\bibitem[{{Hancock} {et~al.}(2013){Hancock}, {Gaensler}, \&
  {Murphy}}]{hancock2013}
{Hancock}, P.~J., {Gaensler}, B.~M., \& {Murphy}, T. 2013, \apj, 776, 106

\bibitem[{{Johnston} {et~al.}(2007){Johnston}, {Bailes}, {Bartel}, {Baugh},
  {Bietenholz}, {Blake}, {Braun}, {Brown}, {Chatterjee}, {Darling}, {Deller},
  {Dodson}, {Edwards}, {Ekers}, {Ellingsen}, {Feain}, {Gaensler}, {Haverkorn},
  {Hobbs}, {Hopkins}, {Jackson}, {James}, {Joncas}, {Kaspi}, {Kilborn},
  {Koribalski}, {Kothes}, {Landecker}, {Lenc}, {Lovell}, {Macquart},
  {Manchester}, {Matthews}, {McClure-Griffiths}, {Norris}, {Pen}, {Phillips},
  {Power}, {Protheroe}, {Sadler}, {Schmidt}, {Stairs}, {Staveley-Smith},
  {Stil}, {Taylor}, {Tingay}, {Tzioumis}, {Walker}, {Wall}, \&
  {Wolleben}}]{johnston07}
{Johnston}, S., {et~al.} 2007, \pasa, 24, 174

\bibitem[{{Klebesadel} {et~al.}(1973){Klebesadel}, {Strong}, \&
  {Olson}}]{klebesadel1973}
{Klebesadel}, R.~W., {Strong}, I.~B., \& {Olson}, R.~A. 1973, ApJ, 182, L85

\bibitem[{{Kouveliotou} {et~al.}(1993){Kouveliotou}, {Meegan}, {Fishman},
  {Bhat}, {Briggs}, {Koshut}, {Paciesas}, \& {Pendleton}}]{kouveliotou1993}
{Kouveliotou}, C., {Meegan}, C.~A., {Fishman}, G.~J., {Bhat}, N.~P., {Briggs},
  M.~S., {Koshut}, T.~M., {Paciesas}, W.~S., \& {Pendleton}, G.~N. 1993, \apjl,
  413, L101

\bibitem[{{Kulkarni} {et~al.}(1999){Kulkarni}, {Frail}, {Sari},
  {Moriarty-Schieven}, {Shepherd}, {Udomprasert}, {Readhead}, {Bloom},
  {Feroci}, \& {Costa}}]{kulkarni1999}
{Kulkarni}, S.~R., {et~al.} 1999, \apjl, 522, L97

\bibitem[{{Laskar} {et~al.}(2013){Laskar}, {Berger}, {Zauderer}, {Margutti},
  {Soderberg}, {Chakraborti}, {Lunnan}, {Chornock}, {Chandra}, \&
  {Ray}}]{laskar2013}
{Laskar}, T., {et~al.} 2013, \apj, 776, 119

\bibitem[{{Levinson} {et~al.}(2002){Levinson}, {Ofek}, {Waxman}, \&
  {Gal-Yam}}]{levinson02}
{Levinson}, A., {Ofek}, E.~O., {Waxman}, E., \& {Gal-Yam}, A. 2002, \apj, 576,
  923

\bibitem[{{Livio} \& {Waxman}(2000)}]{livio00}
{Livio}, M., \& {Waxman}, E. 2000, \apj, 538, 187

\bibitem[{{Lorimer} {et~al.}(2007){Lorimer}, {Bailes}, {McLaughlin},
  {Narkevic}, \& {Crawford}}]{lorimer2007}
{Lorimer}, D.~R., {Bailes}, M., {McLaughlin}, M.~A., {Narkevic}, D.~J., \&
  {Crawford}, F. 2007, Science, 318, 777

\bibitem[{{Malacrino} {et~al.}(2007){Malacrino}, {Atteia}, {Bo{\"e}r}, {Klotz},
  {Veillet}, {Cuillandre}, \& {GRB Rtas Collaboration}}]{malacrino07}
{Malacrino}, F., {Atteia}, J.-L., {Bo{\"e}r}, M., {Klotz}, A., {Veillet}, C.,
  {Cuillandre}, J.-C., \& {GRB Rtas Collaboration}. 2007, \aap, 464, L29

\bibitem[{{Maselli} {et~al.}(2014){Maselli}, {Melandri}, {Nava}, {Mundell},
  {Kawai}, {Campana}, {Covino}, {Cummings}, {Cusumano}, {Evans}, {Ghirlanda},
  {Ghisellini}, {Guidorzi}, {Kobayashi}, {Kuin}, {La Parola}, {Mangano},
  {Oates}, {Sakamoto}, {Serino}, {Virgili}, {Zhang}, {Barthelmy}, {Beardmore},
  {Bernardini}, {Bersier}, {Burrows}, {Calderone}, {Capalbi}, {Chiang},
  {D'Avanzo}, {D'Elia}, {De Pasquale}, {Fugazza}, {Gehrels}, {Gomboc},
  {Harrison}, {Hanayama}, {Japelj}, {Kennea}, {Kopac}, {Kouveliotou}, {Kuroda},
  {Levan}, {Malesani}, {Marshall}, {Nousek}, {O'Brien}, {Osborne}, {Pagani},
  {Page}, {Page}, {Perri}, {Pritchard}, {Romano}, {Saito}, {Sbarufatti},
  {Salvaterra}, {Steele}, {Tanvir}, {Vianello}, {Weigand}, {Wiersema}, {Yatsu},
  {Yoshii}, \& {Tagliaferri}}]{maselli2014}
{Maselli}, A., {et~al.} 2014, Science, 343, 48

\bibitem[{{Meegan} {et~al.}(1992){Meegan}, {Fishman}, {Wilson}, {Horack},
  {Brock}, {Paciesas}, {Pendleton}, \& {Kouveliotou}}]{meegan1992}
{Meegan}, C.~A., {Fishman}, G.~J., {Wilson}, R.~B., {Horack}, J.~M., {Brock},
  M.~N., {Paciesas}, W.~S., {Pendleton}, G.~N., \& {Kouveliotou}, C. 1992,
  \nat, 355, 143

\bibitem[{{Melandri} {et~al.}(2014){Melandri}, {Covino}, {Rogantini},
  {Salvaterra}, {Sbarufatti}, {Bernardini}, {Campana}, {D'Avanzo}, {D'Elia},
  {Fugazza}, {Ghirlanda}, {Ghisellini}, {Nava}, {Vergani}, \&
  {Tagliaferri}}]{melandri14}
{Melandri}, A., {et~al.} 2014, \aap, 565, A72

\bibitem[{{Mesler} \& {Pihlstr{\"o}m}(2013)}]{mesler2013}
{Mesler}, R.~A., \& {Pihlstr{\"o}m}, Y.~M. 2013, \apj, 774, 77

\bibitem[{{Mesler} {et~al.}(2012){Mesler}, {Pihlstr{\"o}m}, {Taylor}, \&
  {Granot}}]{mesler2012}
{Mesler}, R.~A., {Pihlstr{\"o}m}, Y.~M., {Taylor}, G.~B., \& {Granot}, J. 2012,
  \apj, 759, 4

\bibitem[{{Metzger} {et~al.}(1997){Metzger}, {Djorgovski}, {Kulkarni},
  {Steidel}, {Adelberger}, {Frail}, {Costa}, \& {Frontera}}]{metzger1997}
{Metzger}, M.~R., {Djorgovski}, S.~G., {Kulkarni}, S.~R., {Steidel}, C.~C.,
  {Adelberger}, K.~L., {Frail}, D.~A., {Costa}, E., \& {Frontera}, F. 1997,
  \nat, 387, 878

\bibitem[{{Moortgat} \& {Kuijpers}(2006)}]{moortgat2006}
{Moortgat}, J., \& {Kuijpers}, J. 2006, \mnras, 368, 1110

\bibitem[{{Mundell} {et~al.}(2013){Mundell}, {Kopa{\v c}}, {Arnold}, {Steele},
  {Gomboc}, {Kobayashi}, {Harrison}, {Smith}, {Guidorzi}, {Virgili},
  {Melandri}, \& {Japelj}}]{mundell2013}
{Mundell}, C.~G., {et~al.} 2013, \nat, 504, 119

\bibitem[{{Murphy} {et~al.}(2013){Murphy}, {Chatterjee}, {Kaplan}, {Banyer},
  {Bell}, {Bignall}, {Bower}, {Cameron}, {Coward}, {Cordes}, {Croft}, {Curran},
  {Djorgovski}, {Farrell}, {Frail}, {Gaensler}, {Galloway}, {Gendre}, {Green},
  {Hancock}, {Johnston}, {Kamble}, {Law}, {Lazio}, {Lo}, {Macquart}, {Rea},
  {Rebbapragada}, {Reynolds}, {Ryder}, {Schmidt}, {Soria}, {Stairs}, {Tingay},
  {Torkelsson}, {Wagstaff}, {Walker}, {Wayth}, \& {Williams}}]{murphy13}
{Murphy}, T., {et~al.} 2013, \pasa, 30, 6

\bibitem[{{Narayan} {et~al.}(1992){Narayan}, {Paczynski}, \&
  {Piran}}]{narayan1992}
{Narayan}, R., {Paczynski}, B., \& {Piran}, T. 1992, \apjl, 395, L83

\bibitem[{{Nousek} {et~al.}(2006){Nousek}, {Kouveliotou}, {Grupe}, {Page},
  {Granot}, {Ramirez-Ruiz}, {Patel}, {Burrows}, {Mangano}, {Barthelmy},
  {Beardmore}, {Campana}, {Capalbi}, {Chincarini}, {Cusumano}, {Falcone},
  {Gehrels}, {Giommi}, {Goad}, {Godet}, {Hurkett}, {Kennea}, {Moretti},
  {O'Brien}, {Osborne}, {Romano}, {Tagliaferri}, \& {Wells}}]{nousek2006}
{Nousek}, J.~A., {et~al.} 2006, \apj, 642, 389

\bibitem[{{Obenberger} {et~al.}(2014){Obenberger}, {Hartman}, {Taylor},
  {Craig}, {Dowell}, {Helmboldt}, {Henning}, {Schinzel}, \&
  {Wilson}}]{obenberger2014}
{Obenberger}, K.~S., {et~al.} 2014, \apj, 785, 27

\bibitem[{{Paczynski}(2001)}]{paczynski2001}
{Paczynski}, B. 2001, \actaa, 51, 1

\bibitem[{{Perley} \& {Perley}(2013)}]{perley13}
{Perley}, D.~A., \& {Perley}, R.~A. 2013, \apj, 778, 172

\bibitem[{{Perley} {et~al.}(2009){Perley}, {Cenko}, {Bloom}, {Chen}, {Butler},
  {Kocevski}, {Prochaska}, {Brodwin}, {Glazebrook}, {Kasliwal}, {Kulkarni},
  {Lopez}, {Ofek}, {Pettini}, {Soderberg}, \& {Starr}}]{perley2009}
{Perley}, D.~A., {et~al.} 2009, \aj, 138, 1690

\bibitem[{{Perley} {et~al.}(2014){Perley}, {Cenko}, {Corsi}, {Tanvir}, {Levan},
  {Kann}, {Sonbas}, {Wiersema}, {Zheng}, {Zhao}, {Bai}, {Bremer},
  {Castro-Tirado}, {Chang}, {Clubb}, {Frail}, {Fruchter}, {G{\"o}{\u g}{\"u}{\c
  s}}, {Greiner}, {G{\"u}ver}, {Horesh}, {Filippenko}, {Klose}, {Mao},
  {Morgan}, {Pozanenko}, {Schmidl}, {Stecklum}, {Tanga}, {Volnova}, {Volvach},
  {Wang}, {Winters}, \& {Xin}}]{perley14}
---. 2014, \apj, 781, 37

\bibitem[{{Piran}(2004)}]{piran04}
{Piran}, T. 2004, Reviews of Modern Physics, 76, 1143

\bibitem[{{Racusin} {et~al.}(2008){Racusin}, {Karpov}, {Sokolowski}, {Granot},
  {Wu}, {Pal'Shin}, {Covino}, {van der Horst}, {Oates}, {Schady}, {Smith},
  {Cummings}, {Starling}, {Piotrowski}, {Zhang}, {Evans}, {Holland}, {Malek},
  {Page}, {Vetere}, {Margutti}, {Guidorzi}, {Kamble}, {Curran}, {Beardmore},
  {Kouveliotou}, {Mankiewicz}, {Melandri}, {O'Brien}, {Page}, {Piran},
  {Tanvir}, {Wrochna}, {Aptekar}, {Barthelmy}, {Bartolini}, {Beskin}, {Bondar},
  {Bremer}, {Campana}, {Castro-Tirado}, {Cucchiara}, {Cwiok}, {D'Avanzo},
  {D'Elia}, {Della Valle}, {de Ugarte Postigo}, {Dominik}, {Falcone}, {Fiore},
  {Fox}, {Frederiks}, {Fruchter}, {Fugazza}, {Garrett}, {Gehrels},
  {Golenetskii}, {Gomboc}, {Gorosabel}, {Greco}, {Guarnieri}, {Immler},
  {Jelinek}, {Kasprowicz}, {La Parola}, {Levan}, {Mangano}, {Mazets},
  {Molinari}, {Moretti}, {Nawrocki}, {Oleynik}, {Osborne}, {Pagani}, {Pandey},
  {Paragi}, {Perri}, {Piccioni}, {Ramirez-Ruiz}, {Roming}, {Steele}, {Strom},
  {Testa}, {Tosti}, {Ulanov}, {Wiersema}, {Wijers}, {Winters}, {Zarnecki},
  {Zerbi}, {M{\'e}sz{\'a}ros}, {Chincarini}, \& {Burrows}}]{racusin2008}
{Racusin}, J.~L., {et~al.} 2008, \nat, 455, 183

\bibitem[{{Rau} {et~al.}(2006){Rau}, {Greiner}, \& {Schwarz}}]{rau06}
{Rau}, A., {Greiner}, J., \& {Schwarz}, R. 2006, \aap, 449, 79

\bibitem[{{Rhoads}(1997)}]{rhoads1997}
{Rhoads}, J.~E. 1997, \apjl, 487, L1

\bibitem[{{Rol} {et~al.}(2007){Rol}, {van der Horst}, {Wiersema}, {Patel},
  {Levan}, {Nysewander}, {Kouveliotou}, {Wijers}, {Tanvir}, {Reichart},
  {Fruchter}, {Graham}, {Ovaldsen}, {Jaunsen}, {Jonker}, {van Ham}, {Hjorth},
  {Starling}, {O'Brien}, {Fynbo}, {Burrows}, \& {Strom}}]{rol2007}
{Rol}, E., {et~al.} 2007, \apj, 669, 1098

\bibitem[{{Sagiv} \& {Waxman}(2002)}]{sagiv2002}
{Sagiv}, A., \& {Waxman}, E. 2002, \apj, 574, 861

\bibitem[{{Salvaterra} {et~al.}(2012){Salvaterra}, {Campana}, {Vergani},
  {Covino}, {D'Avanzo}, {Fugazza}, {Ghirlanda}, {Ghisellini}, {Melandri},
  {Nava}, {Sbarufatti}, {Flores}, {Piranomonte}, \&
  {Tagliaferri}}]{salvaterra12}
{Salvaterra}, R., {et~al.} 2012, \apj, 749, 68

\bibitem[{{Sari} \& {Piran}(1999)}]{sarip1999}
{Sari}, R., \& {Piran}, T. 1999, \aaps, 138, 537

\bibitem[{{Sari} {et~al.}(1999){Sari}, {Piran}, \& {Halpern}}]{sariph1999}
{Sari}, R., {Piran}, T., \& {Halpern}, J.~P. 1999, \apjl, 519, L17

\bibitem[{{Sari} {et~al.}(1998){Sari}, {Piran}, \& {Narayan}}]{sari1998}
{Sari}, R., {Piran}, T., \& {Narayan}, R. 1998, \apjl, 497, L17

\bibitem[{{Shivvers} \& {Berger}(2011)}]{shivvers11}
{Shivvers}, I., \& {Berger}, E. 2011, \apj, 734, 58

\bibitem[{{Sironi} \& {Giannios}(2013)}]{sironi13}
{Sironi}, L., \& {Giannios}, D. 2013, \apj, 778, 107

\bibitem[{{Soderberg} {et~al.}(2006{\natexlab{a}}){Soderberg}, {Nakar},
  {Berger}, \& {Kulkarni}}]{soderberg2006}
{Soderberg}, A.~M., {Nakar}, E., {Berger}, E., \& {Kulkarni}, S.~R.
  2006{\natexlab{a}}, \apj, 638, 930

\bibitem[{{Soderberg} {et~al.}(2006{\natexlab{b}}){Soderberg}, {Berger},
  {Kasliwal}, {Frail}, {Price}, {Schmidt}, {Kulkarni}, {Fox}, {Cenko},
  {Gal-Yam}, {Nakar}, \& {Roth}}]{soderberg06}
{Soderberg}, A.~M., {et~al.} 2006{\natexlab{b}}, \apj, 650, 261

\bibitem[{{Stanway} {et~al.}(2014){Stanway}, {Levan}, \& {Davies}}]{stanway14}
{Stanway}, E.~R., {Levan}, A.~J., \& {Davies}, L.~J.~M. 2014, \mnras, 444, 2133

\bibitem[{{Tanvir} {et~al.}(2013){Tanvir}, {Levan}, {Fruchter}, {Hjorth},
  {Hounsell}, {Wiersema}, \& {Tunnicliffe}}]{tanvir2013}
{Tanvir}, N.~R., {Levan}, A.~J., {Fruchter}, A.~S., {Hjorth}, J., {Hounsell},
  R.~A., {Wiersema}, K., \& {Tunnicliffe}, R.~L. 2013, \nat, 500, 547

\bibitem[{{Tanvir} {et~al.}(2009){Tanvir}, {Fox}, {Levan}, {Berger},
  {Wiersema}, {Fynbo}, {Cucchiara}, {Kr{\"u}hler}, {Gehrels}, {Bloom},
  {Greiner}, {Evans}, {Rol}, {Olivares}, {Hjorth}, {Jakobsson}, {Farihi},
  {Willingale}, {Starling}, {Cenko}, {Perley}, {Maund}, {Duke}, {Wijers},
  {Adamson}, {Allan}, {Bremer}, {Burrows}, {Castro-Tirado}, {Cavanagh}, {de
  Ugarte Postigo}, {Dopita}, {Fatkhullin}, {Fruchter}, {Foley}, {Gorosabel},
  {Kennea}, {Kerr}, {Klose}, {Krimm}, {Komarova}, {Kulkarni}, {Moskvitin},
  {Mundell}, {Naylor}, {Page}, {Penprase}, {Perri}, {Podsiadlowski}, {Roth},
  {Rutledge}, {Sakamoto}, {Schady}, {Schmidt}, {Soderberg}, {Sollerman},
  {Stephens}, {Stratta}, {Ukwatta}, {Watson}, {Westra}, {Wold}, \&
  {Wolf}}]{tanvir2009}
{Tanvir}, N.~R., {et~al.} 2009, \nat, 461, 1254

\bibitem[{{Taylor} {et~al.}(2004){Taylor}, {Frail}, {Berger}, \&
  {Kulkarni}}]{taylor2004}
{Taylor}, G.~B., {Frail}, D.~A., {Berger}, E., \& {Kulkarni}, S.~R. 2004,
  \apjl, 609, L1

\bibitem[{{Taylor} {et~al.}(2005){Taylor}, {Momjian}, {Pihlstr{\"o}m}, {Ghosh},
  \& {Salter}}]{taylor2005}
{Taylor}, G.~B., {Momjian}, E., {Pihlstr{\"o}m}, Y., {Ghosh}, T., \& {Salter},
  C. 2005, \apj, 622, 986

\bibitem[{{Thornton} {et~al.}(2013){Thornton}, {Stappers}, {Bailes},
  {Barsdell}, {Bates}, {Bhat}, {Burgay}, {Burke-Spolaor}, {Champion}, {Coster},
  {D'Amico}, {Jameson}, {Johnston}, {Keith}, {Kramer}, {Levin}, {Milia}, {Ng},
  {Possenti}, \& {van Straten}}]{thornton2013}
{Thornton}, D., {et~al.} 2013, Science, 341, 53

\bibitem[{{Usov} \& {Katz}(2000)}]{usov2000}
{Usov}, V.~V., \& {Katz}, J.~I. 2000, \aap, 364, 655

\bibitem[{{van der Horst} {et~al.}(2005){van der Horst}, {Rol}, {Wijers},
  {Strom}, {Kaper}, \& {Kouveliotou}}]{vanderhorst05}
{van der Horst}, A.~J., {Rol}, E., {Wijers}, R.~A.~M.~J., {Strom}, R., {Kaper},
  L., \& {Kouveliotou}, C. 2005, \apj, 634, 1166

\bibitem[{{van der Horst} {et~al.}(2008){van der Horst}, {Kamble}, {Resmi},
  {Wijers}, {Bhattacharya}, {Scheers}, {Rol}, {Strom}, {Kouveliotou},
  {Oosterloo}, \& {Ishwara-Chandra}}]{vanderhorst08}
{van der Horst}, A.~J., {et~al.} 2008, \aap, 480, 35

\bibitem[{{van der Horst} {et~al.}(2014){van der Horst}, {Paragi}, {de Bruyn},
  {Granot}, {Kouveliotou}, {Wiersema}, {Starling}, {Curran}, {Wijers},
  {Rowlinson}, {Anderson}, {Fender}, {Yang}, \& {Strom}}]{vanderhorst14}
---. 2014, \mnras, 444, 3151

\bibitem[{{van Eerten} {et~al.}(2012){van Eerten}, {van der Horst}, \&
  {MacFadyen}}]{vaneerten12}
{van Eerten}, H., {van der Horst}, A., \& {MacFadyen}, A. 2012, \apj, 749, 44

\bibitem[{{van Eerten} \& {MacFadyen}(2011)}]{vaneerten11}
{van Eerten}, H.~J., \& {MacFadyen}, A.~I. 2011, \apjl, 733, L37

\bibitem[{{van Paradijs} {et~al.}(2000){van Paradijs}, {Kouveliotou}, \&
  {Wijers}}]{vanparadijs2000}
{van Paradijs}, J., {Kouveliotou}, C., \& {Wijers}, R.~A.~M.~J. 2000, \araa,
  38, 379

\bibitem[{{van Paradijs} {et~al.}(1997){van Paradijs}, {Groot}, {Galama},
  {Kouveliotou}, {Strom}, {Telting}, {Rutten}, {Fishman}, {Meegan}, {Pettini},
  {Tanvir}, {Bloom}, {Pedersen}, {N{\o}rdgaard-Nielsen}, {Linden-V{\o}rnle},
  {Melnick}, {van der Steene}, {Bremer}, {Naber}, {Heise}, {in't Zand},
  {Costa}, {Feroci}, {Piro}, {Frontera}, {Zavattini}, {Nicastro}, {Palazzi},
  {Bennett}, {Hanlon}, \& {Parmar}}]{vanparadijs1997}
{van Paradijs}, J., {et~al.} 1997, \nat, 386, 686

\bibitem[{{Wiersema} {et~al.}(2014){Wiersema}, {Covino}, {Toma}, {van der
  Horst}, {Varela}, {Min}, {Greiner}, {Starling}, {Tanvir}, {Wijers},
  {Campana}, {Curran}, {Fan}, {Fynbo}, {Gorosabel}, {Gomboc}, {Gotz}, {Hjorth},
  {Jin}, {Kobayashi}, {Kouveliotou}, {Mundell}, {O'Brien}, {Pian}, {Rowlinson},
  {Russell}, {Salvaterra}, {di Serego Alighieri}, {Tagliaferri}, {Vergani},
  {Elliott}, {Farina}, {Hartoog}, {Karjalainen}, {Klose}, {Knust}, {Levan},
  {Schady}, {Sudilovsky}, \& {Willingale}}]{wiersema2014}
{Wiersema}, K., {et~al.} 2014, \nat, 509, 201

\bibitem[{{Wijers} \& {Galama}(1999)}]{wijersg1999}
{Wijers}, R.~A.~M.~J., \& {Galama}, T.~J. 1999, \apj, 523, 177

\bibitem[{{Wijers} {et~al.}(1999){Wijers}, {Vreeswijk}, {Galama}, {Rol}, {van
  Paradijs}, {Kouveliotou}, {Giblin}, {Masetti}, {Palazzi}, {Pian}, {Frontera},
  {Nicastro}, {Falomo}, {Soffitta}, \& {Piro}}]{wijers1999}
{Wijers}, R.~A.~M.~J., {et~al.} 1999, \apjl, 523, L33

\bibitem[{{Woosley}(1993)}]{woosley1993}
{Woosley}, S.~E. 1993, \apj, 405, 273

\bibitem[{{Woosley} \& {Bloom}(2006)}]{woosley2006}
{Woosley}, S.~E., \& {Bloom}, J.~S. 2006, \araa, 44, 507

\bibitem[{{Wygoda} {et~al.}(2011){Wygoda}, {Waxman}, \& {Frail}}]{wygoda11}
{Wygoda}, N., {Waxman}, E., \& {Frail}, D.~A. 2011, \apjl, 738, L23

\bibitem[{{Zauderer} {et~al.}(2013){Zauderer}, {Berger}, {Margutti}, {Levan},
  {Olivares E.}, {Perley}, {Fong}, {Horesh}, {Updike}, {Greiner}, {Tanvir},
  {Laskar}, {Chornock}, {Soderberg}, {Menten}, {Nakar}, {Carpenter}, {Chandra},
  {Castro-Tirado}, {Bremer}, {Gorosabel}, {Guziy}, {P{\'e}rez-Ram{\'{\i}}rez},
  \& {Winters}}]{zauderer2013}
{Zauderer}, B.~A., {et~al.} 2013, \apj, 767, 161

\bibitem[{{Zhang}(2014)}]{zhang2014}
{Zhang}, B. 2014, \apjl, 780, L21

\bibitem[{{Zhang} \& {MacFadyen}(2009)}]{zhang09}
{Zhang}, W., \& {MacFadyen}, A. 2009, \apj, 698, 1261

\bibitem[{{Zou} {et~al.}(2007){Zou}, {Wu}, \& {Dai}}]{zou07}
{Zou}, Y.~C., {Wu}, X.~F., \& {Dai}, Z.~G. 2007, \aap, 461, 115

\end{thebibliography}

\end{document}